\documentclass[superscriptaddress,aps,preprintnumbers,amsmath,amssymb,prd,nofootinbib,preprint,longbibliography]{revtex4-1}
\pdfoutput=1

\usepackage{cancel}
\usepackage{graphicx,array}
\usepackage{color}
\usepackage{latexsym}
\usepackage{amsthm}
\usepackage{amsmath}
\usepackage{amssymb}
\usepackage{hyperref} 
\usepackage{bbold}
\usepackage{mathtools}
\usepackage{enumitem}
\usepackage{makecell}
\usepackage[table]{xcolor}
\usepackage{multirow}
\usepackage{booktabs}
\usepackage{setspace}
\usepackage{stackengine}
\usepackage{comment}

\numberwithin{equation}{section}

\newcommand\barparen[1]{\overset{(-)}{#1}}

\makeatletter
\renewcommand{\p@subsection}{}
\renewcommand{\p@subsubsection}{}
\makeatother

\def\simgt{\mathrel{\lower2.5pt\vbox{\lineskip=0pt\baselineskip=0pt
           \hbox{$>$}\hbox{$\sim$}}}}
\def\simlt{\mathrel{\lower2.5pt\vbox{\lineskip=0pt\baselineskip=0pt
           \hbox{$<$}\hbox{$\sim$}}}}

\newcommand{\be}{\begin{equation}}
\newcommand{\ee}{\end{equation}}
\newcommand{\bea}{\begin{eqnarray}}
\newcommand{\eea}{\end{eqnarray}}

\newcommand{\eV}{\textrm{ eV}}

\newcommand{\MeV}{\textrm{ MeV}}
\newcommand{\GeV}{\textrm{ GeV}}
\newcommand{\TeV}{\textrm{ TeV}}
\newcommand{\gsim}{\lower.7ex\hbox{$\;\stackrel{\textstyle>}{\sim}\;$}}
\newcommand{\lsim}{\lower.7ex\hbox{$\;\stackrel{\textstyle<}{\sim}\;$}}

\newcommand{\MPl}{M_{\rm Pl}}

\definecolor{nicered}{rgb}{0.7,0.1,0.1}
\definecolor{nicegreen}{rgb}{0.1,0.5,0.1}
\definecolor{PatColor}{rgb}{0,.8,0}
\hypersetup{colorlinks,citecolor= blue,linkcolor= black}

\begin{document}

\title{Lepto-axiogenesis with light right-handed neutrinos}

\author{Patrick Barnes}
\affiliation{Leinweber Center for Theoretical Physics, Department of Physics, University of Michigan, Ann Arbor, MI 48109, USA}
\author{Raymond T. Co}
\affiliation{William I. Fine Theoretical Physics Institute, School of Physics and Astronomy, University of Minnesota, Minneapolis, MN 55455, USA}
\affiliation{Physics Department, Indiana University, Bloomington, IN, 47405, USA}
\author{Keisuke Harigaya}
\affiliation{Department of Physics, University of Chicago, Chicago, IL 60637, USA}
\affiliation{Enrico Fermi Institute and Kavli Institute for Cosmological Physics, University of Chicago, Chicago, IL 60637, USA}
\affiliation{Kavli Institute for the Physics and Mathematics of the Universe (WPI),\\
        The University of Tokyo Institutes for Advanced Study,\\
        The University of Tokyo, Kashiwa, Chiba 277-8583, Japan}
\author{Aaron Pierce}
\affiliation{Leinweber Center for Theoretical Physics, Department of Physics, University of Michigan, Ann Arbor, MI 48109, USA}

\date{\today}

\begin{abstract}
We study lepto-axiogenesis in theories where the right-handed neutrino is light enough that its dynamics affect the determination of the baryon asymmetry. When compared with theories of high-scale lepto-axiogenesis where the Majorana neutrino mass may be treated as an effective dimension-five operator, we find that the predicted saxion mass is lower.   Two distinct scenarios emerge.  In the first, processes that generate the baryon asymmetry are in equilibrium down to the mass of the right-handed neutrino.  In the second, the relevant processes never reach equilibrium; the baryon number freezes in. We comment on implications for supersymmetric spectra and discuss constraints on late decays of supersymmetric relics and from dark radiation. In contrast to high-scale lepto-axiogenesis, which predicts superpartners with masses of 10-100 TeV or more, we find this scenario is consistent with a wider range of superpartner masses, all the way down to current direct search bounds. 
\end{abstract}

\preprint{LCTP-24-02, UMN-TH-4307/23, FTPI-MINN-23-27}

\maketitle

\vspace{1cm}

\begingroup
\hypersetup{linkcolor=black}
\renewcommand{\baselinestretch}{1.26}\normalsize
\tableofcontents
\renewcommand{\baselinestretch}{2}\normalsize
\endgroup

\newpage

\section{Introduction}
\label{sec:intro}

The axion solves the strong CP problem~\cite{Peccei:1977hh, Peccei:1977ur,Weinberg:1977ma,Wilczek:1977pj} and is a dark matter candidate~\cite{Preskill:1982cy, Dine:1982ah,Abbott:1982af}. Furthermore, it may be responsible for the observed baryon asymmetry of the universe.  The axion may rotate in field space in the early universe as in the Affleck-Dine mechanism~\cite{Affleck:1984fy,Dine:1995kz}, and the angular momentum of the axion field, which corresponds to a non-zero Peccei-Quinn (PQ) charge, may be transformed into a baryon asymmetry~\cite{Co:2019wyp,Co:2020jtv,Harigaya:2021txz,Chakraborty:2021fkp,Kawamura:2021xpu,Co:2021qgl,Co:2022aav,Barnes:2022ren,Co:2022kul,Badziak:2023fsc}; see also related work using other pseudo-Nambu Goldstone bosons in Refs.~\cite{Co:2020xlh,Berbig:2023uzs,Chao:2023ojl,Chun:2023eqc}. 

In minimal axiogenesis~\cite{Co:2019wyp}, the PQ charge is transferred to a baryon asymmetry via Standard Model sphaleron processes.  However, this minimal model underproduces the baryon abundance once the axion dark matter abundance is fixed by the kinetic misalignment mechanism (KMM).  In the KMM~\cite{Co:2019jts}, the axion's non-zero velocity in field space impacts dark matter production; it allows for the generation of the observed dark matter density for lower axion decay constants than the conventional misalignment mechanism~\cite{Preskill:1982cy, Dine:1982ah,Abbott:1982af}. The underproduction of baryon number in  minimal axiogenesis can be remedied by changing the electroweak phase transition temperature~\cite{Co:2019wyp} or taking a small decay constant $f_a < 10^7$ GeV while suppressing axion couplings relevant for astrophysical constraints~\cite{Badziak:2023fsc}.
The baryon asymmetry can be also generated by exponential production of helical hypercharge magnetic fields by tachyonic instability, but fine-tuning of the parameters is required~\cite{Co:2022kul}. 

Alternatively, there can be new physics that increases the efficiency of the PQ charge transfer.  This new physics should generate baryon number minus lepton number ($B-L$) so that the additional baryon number is immune to washout by electroweak sphaleron processes.  A well-motivated and simple possibility takes advantage of neutrino masses generated by lepton number-violating Majorana masses \cite{Domcke:2020kcp,Co:2020jtv,Kawamura:2021xpu,Barnes:2022ren}.  This has previously been investigated under the assumption that the right-handed neutrinos are heavy, with masses above all the relevant temperatures for this cosmology.   We refer to this possibility as ``high-scale lepto-axiogenesis."  In high-scale lepto-axiogenesis, the additional $B-L$ violation is provided by the effective dimension-five operator $(\ell H)^2$.  In this work, we explore the possibility that right-handed neutrinos are light enough that the description in terms of this effective operator fails. This impacts the efficiency of the baryon asymmetry generation and hence the allowed parameter space.

We focus on the ramifications for supersymmetric theories, where axiogenesis is realized most straightforwardly.  Supersymmetric theories can naturally have an approximately flat direction for the PQ field.  This allows the radial direction of the PQ field to take on a large field value following inflation~\cite{Dine:1995kz}. This, in turn, enhances the effects of higher-dimensional PQ-violating operators. These operators provide a ``kick" for the PQ field, allowing it to rotate in field space.  This angular momentum in field space corresponds to the non-zero PQ charge that is eventually converted to baryon number.

Reproducing the observed baryon symmetry (and dark matter abundance) via the lepto-axiogenesis mechanism fixes the mass $m_{S}$ of the radial direction of the PQ field, the saxion.  In many models $m_{S}$ is connected to the masses of the superpartners, so fixing the baryon asymmetry allows a prediction of the superpartner masses.  In high-scale lepto-axiogenesis, the scalar superpartner masses are large, at the 10-100 TeV scale~\cite{Co:2020jtv,Barnes:2022ren}. A consequence is that the simplest theories of  weak-scale supersymmetry---TeV-scale gravity-mediation or gauge-mediation---seem to be in tension with lepto-axiogenesis.  Here we explore the case where right-handed neutrino masses $m_N$ are much lower, so that the small observed neutrino masses are explained not just by the seesaw mechanism~\cite{Yanagida:1979as,GellMann:1980vs,Minkowski:1977sc,Mohapatra:1979ia}, but rather a combination of the seesaw mechanism and a small neutrino Yukawa coupling.

We will see that the lower right-handed neutrino masses can qualitatively change the lepto-axiogenesis story. In high-scale lepto-axiogenesis, the dominant contribution to the baryon asymmetry is generated in the era between the end of inflationary reheating and when the saxion reaches the minimum of its potential.  But here, as long as the neutrino Yukawa couplings are sufficiently large, the asymmetry is generated via a freeze-out mechanism, with freeze-out occurring at  $T\sim m_N$.  There is also a second possibility wherein tiny Yukawa couplings allow an asymmetry to be generated via a freeze-in mechanism. 
In both cases, we find $m_S$ can be much below the $10-100$ TeV scale, and so low-scale lepto-axiogenesis can be embedded into a broader class of supersymmetry-breaking and mediation scenarios. 

In this setup there are both relics that are stable and long-lived.  Requiring that these relics be consistent with cosmological data gives important constraints.  Consistency with Big Bang Nucleosynthesis (BBN) \cite{Kawasaki:2008qe,Kawasaki:2017bqm} is particularly important.  As an example, if sufficiently light, the right-handed neutrino can have very small Yukawa couplings and is therefore long-lived. Because we are working in a supersymmetric context, additional relics of interest include the gravitino, the axino, the right-handed sneutrino, and the lightest Minimal Supersymmetric Standard Model (MSSM) superpartner.   


In the next section we provide a brief review of the lepto-axiogenesis scenario.  Then in Sec.~\ref{sec:FreezeOut}, we discuss the case of right-handed neutrinos with masses above the TeV scale.  In this case, we will see that the baryon asymmetry is typically generated via a freeze-out mechanism.   
We then discuss constraints on the spectrum arising from ensuring that potentially long-lived (or stable) relics give a consistent cosmology.
In Sec.~\ref{sec:DL} we discuss the possibility of sub-weak-scale right-handed neutrinos.  Owing to the smaller Yukawa couplings present in this case, the asymmetry is generated via freeze-in.   Additional details on the calculations relevant for the freeze-in scenario are given in Appendix \ref{sec:AppendixFI}.
When the saxion oscillates with a large field value, it can undergo parametric resonance before settling to its minimum.  This can lead to an additional cosmological constraints depending on the shape of the saxion potential, which we discuss in Sec.~\ref{sec:DWandPR}. Finally, in Sec.~\ref{sec:conclusion}, we summarize our findings, and discuss some implications of this work.

\section{Review of lepto-axiogenesis}
\label{sec:review}

In lepto-axiogenesis, the nonzero PQ charge is carried by the rotation in field space of the PQ symmetry-breaking field $P$.  This charge is subsequently converted to baryon number. We will review these two aspects in the following two subsections.

\subsection{Dynamics of the Peccei-Quinn field}
\label{subsec:dynamics}

The PQ symmetry breaking field $P$ contains a radial mode $S$, which we call the saxion, and the axial mode $a$, the axion:
\begin{equation}
    P = \frac{(f_a N_{\rm DW} +S)}{\sqrt{2}} e^{i \frac{a}{ f_a N_{ \rm DW}}}.
\end{equation}
Here, $N_{\rm DW}$ is the domain wall number, and $f_{a}$ is the axion decay constant. It is useful to introduce $\theta \equiv a/f_a$, the angle of the axial mode in field space.  If the field $P$ rotates in field space, i.e., $\dot\theta \neq 0$, a non-zero PQ charge exists and is given by
\begin{align}
\label{eq:chargedensity}
    n_{\theta} =  \frac{i}{N_{\rm DW}} \left(\dot{P}P^*-  \dot{P}^* P \right) 
    = - \dot{\theta} \left(f_a + \frac{S}{N_{\rm DW}}\right)^2.
\end{align}

The vacuum potential from supersymmetry breaking can take one of two forms. One possibility is that the mass of the saxion gets a radiative correction from a Yukawa coupling of the saxion with a PQ-charged fermion, e.g., a KSVZ quark.  In this case, the potential generated via dimensional transmutation takes the form 
\begin{align}
\label{eq:one-field}
V_{\rm soft}(P) = \frac{1}{2} m_S^2 |P|^2 \left( {\rm ln} \frac{2 |P|^2}{f_a^2 N_{\rm DW}^2} -1 \right).
\end{align}
 Alternatively, if there exists a chiral multiplet $X$ whose $F$-term potential creates a moduli space for two PQ-charged fields $P$ and $\bar{P}$, then the potential arises from the superpotential and soft supersymmetry breaking terms
\begin{align}
\label{eq:W_two-field}
    W = \lambda X (P \bar{P} - v_{\rm PQ}^2),~~V_{\rm soft}(P) = m_P^2 |P|^2 + m_{\bar{P}}^2|\bar{P}|^2. 
\end{align}
Without loss of generality, we assume that $|P| \gg |\bar{P}|$ in the early universe. In this case, $\bar{P}$ in Eq.~(\ref{eq:W_two-field}) can be integrated out and replaced by $v_{\rm PQ}^2 / P$.  Then the potential of $P$ becomes
\begin{equation}
\label{eq:two-field}
    V_{\rm soft}(P) \simeq m_P^2 |P^2| \left( 1 + \frac{m_{\bar{P}}^2}{m_P^2} \frac{v_{PQ}^4}{|P^4|} \right) .
\end{equation}

In either case, the initiation of the rotation is analogous to that in the Affleck-Dine baryogenesis mechanism~\cite{Affleck:1984fy,Dine:1995kz}, which we now review in this context. The PQ symmetry may be broken by higher-dimensional operators arising from a superpotential of the form,
\begin{align}
    W_{\cancel{PQ}}=\frac{1}{q}\frac{P^{q}}{M^{q-3}}, 
\end{align} 
with an integer $q$ sufficiently large so as to avoid the axion quality problem \cite{Holman:1992us,Barr:1992qq,Kamionkowski:1992mf,Dine:1992vx}.  Including the PQ-breaking term, the full potential of $P$ at large field values is given by 
\begin{equation}
\label{eq:VAterm}
        V(P) = (m_S^2- c_H H^2)|P|^2 + \frac{|P|^{2q-2}}{M^{2q-6}} + \left( A\frac{P^q}{M^{q-3}} + {\rm h.c.}\right),
\end{equation}
where we have made the approximation that the soft supersymmetry breaking potential is nearly quadratic when $|P| \gg f_a$ according to Eqs.~(\ref{eq:one-field}) and (\ref{eq:two-field}), and $c_{H}$ is the coefficient of the Hubble induced mass \cite{Dine:1995kz} that allows for a large saxion field value following inflation.
This large saxion field value enhances the importance of the higher-dimensional PQ-breaking term.  This term then gives a kick to the $P$ field in the axial direction.  The $P$ field will begin oscillation when $m_{S} \sim H$, and the radial direction will eventually settle to its minimum at a temperature $T_{S}$.  Before $T_{S}$, $\dot{\theta}$ is constant, with a value given by 
\begin{equation}
    \dot\theta = N_{\rm DW} m_S   \hspace{1cm}  {\rm for}~~~T > T_{S} .
\end{equation}
After $T_{S}$, $\dot{\theta}$ redshifts as $T^3$ as can be seen from charge conservation: $n_\theta \propto \dot\theta f_a^2 \propto T^3$. 

 The rotation of the axion impacts the generation of axion dark matter and modifies it from the conventional misalignment case.  This motion in field space delays axion oscillations around the minimum relative to the conventional misalignment picture.  In fact, a process called axion fragmentation~\cite{Jaeckel:2016qjp, Berges:2019dgr, Fonseca:2019ypl} occurs before the rotation transitions into oscillations. This destroys the coherent motion of the axion and creates axion fluctuations.\footnote{However, the final estimation of the axion abundance is largely unaltered from just working under the assumption that the primary effect of the rotation is the delay of oscillations~\cite{Co:2021rhi,Morgante:2021bks,Eroncel:2022vjg}.} Requiring that the dark matter be comprised of axions produced via the kinetic misalignment mechanism fixes the required PQ charge yield~\cite{Co:2019jts}
\begin{align}
\label{eq:KMM_Ytheta}
    Y_\theta \equiv \frac{n_\theta}{s} \simeq \pm\frac{0.44 \eV}{m_a} ,
\end{align}
with $n_\theta$ the charge density, see Eq.~(\ref{eq:chargedensity}), and $s$ the entropy density. This, in turn, allows us to determine the temperature at which the saxion reaches its minimum.  We first assume the saxion reaches its minimum during the radiation-dominated era following the completion of reheating. In this case, $T_{S}$ can be determined by considering the yield at the time the saxion reaches its minimum
\begin{equation}
\frac{\rho_a}{s}
= m_a \left|Y_\theta\right| = m_a \frac{N_{\rm DW} m_S f_a^2}{\frac{2 \pi^2 g_*(T_{S})}{45}T_{S}^3 }.
\label{eq:KMMstep}
\end{equation}
Then noting that the subsequent evolution is adiabatic and using Eq.~(\ref{eq:KMM_Ytheta}), we find
\begin{equation}
\label{eq:TS_KMM}
T_{S,\,{\rm KMM}}^{\rm rad-dom} \simeq 1.6 \times 10^6 \GeV 
\left( \frac{N_{\rm DW}}{3} \right)^{ \scalebox{1.01}{$\frac{1}{3}$} }
\left( \frac{m_S}{10 \TeV} \right)^{ \scalebox{1.01}{$\frac{1}{3}$} } 
\left( \frac{f_a}{10^9 \GeV} \right)^{ \scalebox{1.01}{$\frac{1}{3}$} }
\left(\frac{g_{\rm MSSM}}{g_*(T_S)}\right)^{ \scalebox{1.01}{$\frac{1}{3}$} },
\end{equation}
where $g_{\rm MSSM} = 228.75$ is the effective degrees of freedom for the MSSM.
If the saxion instead reaches its minimum during reheating that proceeds by perturbative decays where $H \propto T^4$, entropy is produced during the reheating process. $T_{S}$ is modified accordingly
\begin{equation}
\label{eq:TS_KMM_TR}
T_{S,\,{\rm KMM}}^{\rm reheating} \simeq 67 \TeV 
\left( \frac{N_{\rm DW}}{3} \right)^{ \scalebox{1.01}{$\frac{1}{8}$} }  
\left( \frac{m_S}{10 \TeV} \right)^{ \scalebox{1.01}{$\frac{1}{8}$} } 
\left( \frac{f_a}{10^9 \GeV} \right)^{ \scalebox{1.01}{$\frac{1}{8}$} }
\left(\frac{g_{\rm MSSM}}{g_*(T_R)}\right)^{ \scalebox{1.01}{$\frac{1}{8}$} }
\left(\frac{T_R}{10 \TeV}\right)^{ \scalebox{1.01}{$\frac{5}{8}$} } .
\end{equation}
This expression is of particular interest if the reheat temperature is low, which could be required by constraints arising from supersymmetric relics such as the gravitino, see Sec.~\ref{sec:cosmologicalrelics}.

The energy density of the rotating field, $\dot\theta Y_\theta s$, which scales as $T^3$ before $T = T_S$, can come to dominate the radiation energy density, $\rho_{\rm rad}=\pi^2 g_* T^4 / 30$, at a temperature
\begin{align}
\label{eq:TRM}
    T_{\rm RM} = \frac{4}{3} N_{\rm DW} m_S \left|Y_\theta\right| \simeq 100 \TeV 
    \left( \frac{m_S}{\rm TeV} \right)
    \left( \frac{f_a}{10^9 \GeV} \right) N_{\rm DW}.
\end{align}
 In the last expression, the charge yield is fixed to the value required by the KMM given in Eq.~(\ref{eq:KMM_Ytheta}). Rotation domination will occur if $T_{\rm RM} > T_S$ because the energy density in the rotation redshifts as $T^6$ when $T < T_S$. When rotation domination does occur, the universe undergoes an era of matter- and kination-domination  before and after $T_S$, respectively.  See Ref.~\cite{Co:2021qgl} for a detailed 
review of the dynamics of the PQ field. Such a rotation-dominated era may imprint a unique triangular peak in a primordial gravitational wave spectrum~\cite{Co:2021lkc,Gouttenoire:2021wzu,Gouttenoire:2021jhk}, initially produced by, e.g., inflation or cosmic strings.

\subsubsection{Thermalization of the Peccei-Quinn Field via PQ-charged matter}
\label{sec:Therm}
As in the Affleck-Dine mechanism, when the rotation commences, it is generically elliptic rather than circular; it is a superposition of an angular motion and a radial motion. The radial motion should be depleted early enough, since a late depletion leads to the domination of the energy density of the Universe by the radial mode and entropy production upon its eventual decay. This can, in principle, excessively dilute the PQ charge. The radial motion may be depleted by interactions with the thermal bath, i.e., thermalization.  

We will find that there are two viable thermal histories.  In the first, the thermalization occurs prior to the radial density coming to dominate, i.e. $T_{\rm th} > T_{\rm RM}$, thereby avoiding the substantial release of entropy upon thermalization.  Note, we have previously defined $T_{\rm RM}$ in  Eq.~\eqref{eq:TRM} as the temperature where the energy density of the \emph{rotation} may come to dominate. The temperature where the energy density of the radial mode (saxion), may come to dominate is similar. Saxion domination would occur at temperature a factor $\epsilon^{-1}$ larger than $T_{\rm RM}$, where $\epsilon^{-1}$ describes the ellipticity of the rotation. In the following we take $\epsilon = 1$ for simplicity. Indeed, $\epsilon \sim 1$ is expected when the parts of the potential that initiate the rotation are dominated by gravity-mediated effects;  then $A$ in Eq.~(\ref{eq:VAterm}) is expected to be of the same order as $m_S$ ($\sim m_{3/2}$), and the potential gradient in the angular direction is of the same order as that in the radial direction. In the second viable cosmology there is a period where the energy density of the radial mode dominates, but the entropy generation is not so large as to preclude the possibility of axion dark matter.  Furthermore, as we will emphasize in the following, this entropy generation can occur prior to the generation of the baryon asymmetry, so that the calculation of the asymmetry is unmodified.  We now turn to the ways in which thermalization may occur.

One possibility is that thermalization proceeds via interactions with PQ-charged particles $Q$ present in the bath with a coupling $y_Q P Q \bar Q$. In the case of the KSVZ model, these fields could be the PQ fermions responsible for the generation of the QCD anomaly and the solution to the strong CP problem.  In the DFSZ case, the existence of such $Q$ would be an extension to the model. The thermalization rate is given by $\Gamma_{\rm th} \simeq b y_Q^2 T$ where $b \simeq 0.1$~\cite{Mukaida:2012qn} assuming that $Q$ has a gauge charge. These $Q$ can be heavy, and when $P$ is at large field value, may have a mass greater than the temperature of the thermal bath and would not be present in large numbers in the bath, which would temporarily prevent thermalization of the saxion. However, as $P$ redshifts, the $Q$ mass decreases, and these PQ-charged particles can come in to equilibrium.  The requirement that $Q$ be in the bath at a given temperature $T$, $y_Q S \lesssim T$, limits the maximum possible thermalization rate $\Gamma_{\rm th} \simeq b y_{Q}^2 T \lesssim b T^3/S^2\equiv \Gamma_{\rm th}^{\rm max}$. Because $S$ scales as $T^{3/2}$ for all temperatures prior to $S$ reaching its minimum ($T> T_{S}$) during a radiation-dominated era, $\Gamma_{\rm th}^{\rm max} = b T_{S}^3/f_{a}^2$ for all such times. But for temperatures below $T_{S}$, the maximum thermalization rate will fall faster than the Hubble rate, so if the thermalization has not occurred by this time even with the maximal thermalization rate, it will not occur. Requiring that thermalization does indeed occur limits the allowed range of $m_{S}$ and $f_{a}$, and by imposing $\Gamma_{\rm th}^{\rm max}(T_S) \ge 3 H(T_S)$ we find
\begin{equation}
\label{eq:thermRadiation}
    f_a \lesssim 
        4 \times 10^{10} \GeV  
        \left( \frac{b}{0.1} \right)^{\scalebox{1.01}{$\frac{3}{5}$} }
        \left( \frac{m_S}{0.1 \GeV} \right)^{\scalebox{1.01}{$\frac{1}{5}$} } 
        \left( \frac{g_{\rm MSSM}}{g_*(T_S)} \right)^{\scalebox{1.01}{$\frac{1}{2}$} } 
        N_{\rm DW}^{1/5} .
\end{equation}
For $T_{\rm RM} < T_S$ as discussed around Eq.~(\ref{eq:TRM}), this constraint automatically ensures that the saxion does not dominate the energy density. 

 If $T_{\rm RM} > T_S$,  the consistency condition to ensure that no entropy is created by the saxion thermalization, is given by $\Gamma_{\rm th}^{\rm max}(T_{\rm RM}) \ge 3 H(T_{\rm RM})$ and reads
\begin{equation}
\label{eq:thermTRM}
    f_a \lesssim 
        6 \times 10^{10} \GeV  
        \left( \frac{b}{0.1} \right)^{\scalebox{1.01}{$\frac{1}{3}$} }
        \left( \frac{\rm TeV}{m_S} \right)^{\scalebox{1.01}{$\frac{1}{3}$} } 
        \left( \frac{g_{\rm MSSM}}{g_*(T_{\rm th})} \right)^{\scalebox{1.01}{$\frac{1}{2}$} } N_{\rm DW}^{-1/3}.
\end{equation}
Should instead the PQ field dominate the energy density at the time of thermalization, entropy is created upon thermalization.   However, a viable cosmology is still possible, as we now discuss. 
 The resulting PQ charge yield is given by $Y_\theta = \rho_{\rm th} / (\dot\theta  s) = 3 T_{\rm th} / 4 \dot\theta$. Because this yield depends on the thermalization temperature, $T_{\rm th}$ must be such that the correct dark matter yield is achieved. Fixing $T_{\rm th}$ to the value required by the dark matter abundance using Eq.~(\ref{eq:KMM_Ytheta}) imposes a constraint equivalent to Eq.~(\ref{eq:thermTRM}) but with the inequality saturated.  This represents a second consistent cosmology.

Finally, we note that if additional PQ-charged particles $Q \bar Q$ are not present, the saxion will scatter with the gluons at the rate $\Gamma_{g} \simeq N_{\rm DW}^2 10^{-5} T^3/S^2$~\cite{Mukaida:2012qn}. In this case, the constraints above will apply but with $b$ set to $N_{\rm DW}^2 10^{-5}$. 

\subsubsection{DFSZ: thermalization of the PQ-field via interactions with Higgs fields}
\label{sec:thermHiggs}

\begin{figure}
\includegraphics[width=0.495\linewidth]{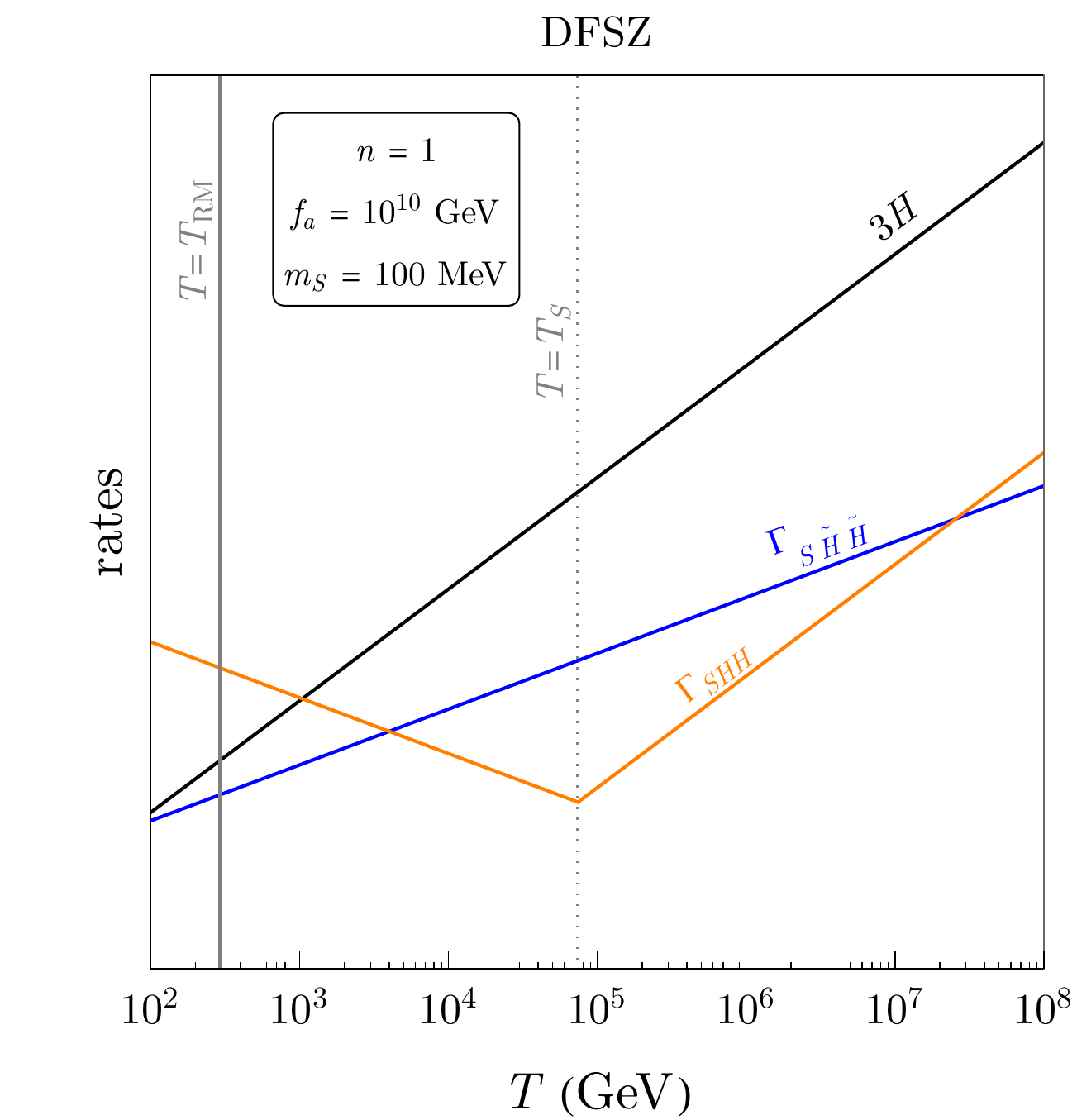}
\includegraphics[width=0.495\linewidth]{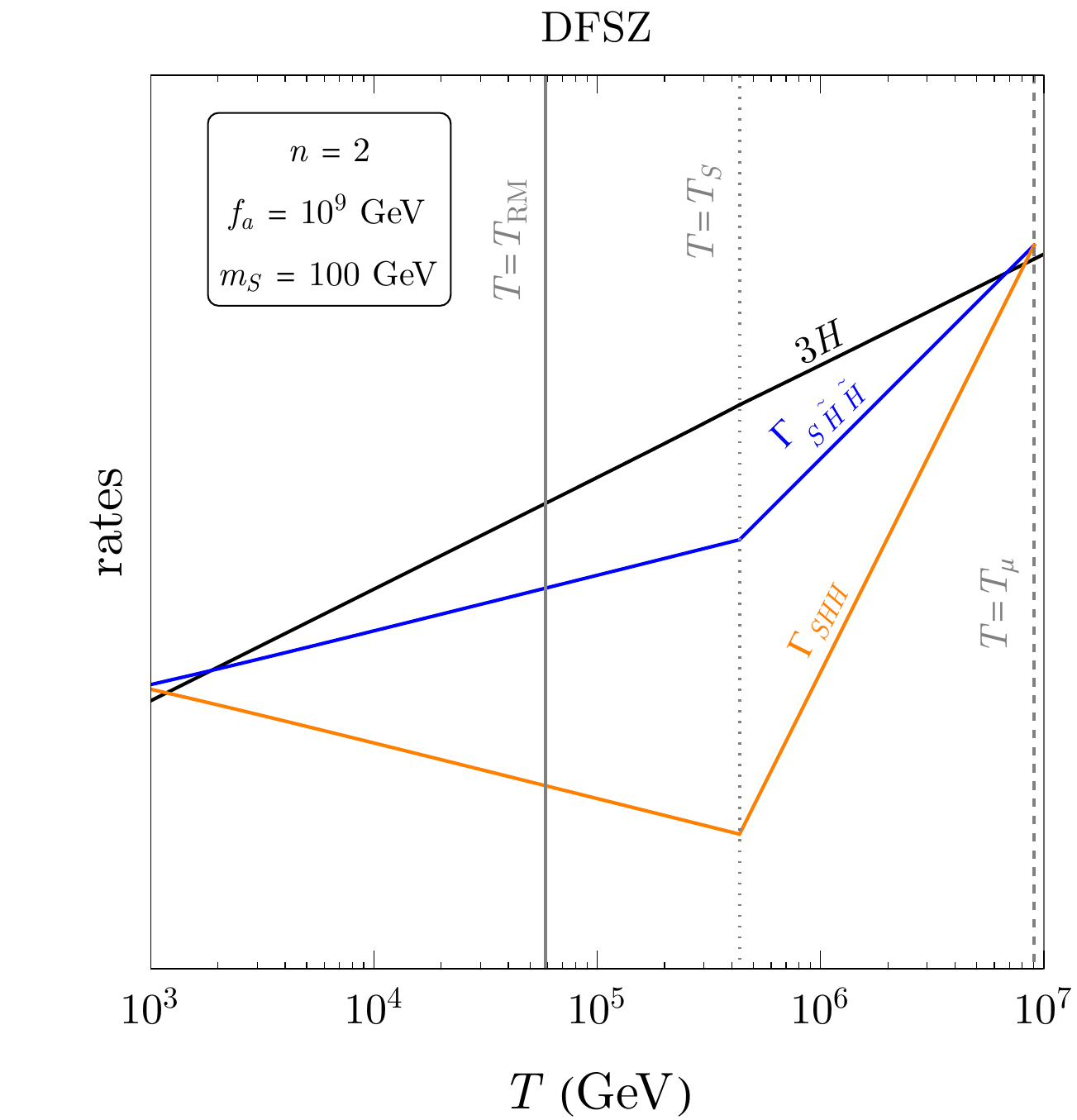}
\caption{The Hubble expansion rate (black) and the thermalization rates for the saxion using interactions with Higgsinos (blue) and Higgs fields (red) in the DFSZ model with $n=1$ (left) and $n=2$ (right) for $\mu = 2 \TeV$ and different benchmark values of $f_a$ and $m_S$ as shown. Also shown is the temperature $T_{S}$ where the saxion reaches its minimum, and $T_{\rm RM}$ where the rotation would come to dominate over the radiation energy density if not thermalized.}
\label{fig:DFSZTherm}
\end{figure}

In the minimal DFSZ model, the PQ-breaking field couples only to the Higgs multiplets
\begin{equation}
\label{eq:DFSZDef}
    W \supset \mu \left( \frac{P}{\frac{N_{\rm DW} f_a }{\sqrt{2} }} \right)^n H_u H_d ,
\end{equation}
where $\mu$ is the supersymmetric mass of the Higgs multiplets, and thermalization proceeds via the corresponding interactions.  Here, we will consider $n=1,2$.  Similar analyses could apply for larger $n$, but as $n$ increases obtaining a sufficiently large $\mu$ would require a relatively low cutoff scale suppressing this higher-dimensional operator.
The interaction rates with the Higgsinos and the Higgs fields are given by~\cite{Barnes:2022ren}
\begin{align}
\label{eq:Gamma_S_higgsino}
\Gamma_{S\widetilde{H}\widetilde{H}} & \simeq 0.1 \frac{n^2\mu^2 (T+m_S)}{S^2} \left(\frac{S}{N_{\rm DW} f_a}\right)^{2n} , \nonumber \\
\Gamma_{SHH} & \simeq  0.1 \frac{4 n^2\mu^4}{S^2(T+m_S)} \left(\frac{S}{N_{\rm DW} f_a}\right)^{4n} ,
\end{align}
where high/low temperature limits correspond to the scattering/decay processes. For $n=1$, these thermalization rates either scale the same way or decrease more slowly than the Hubble expansion for $T>T_S$ as illustrated in the left panel of Fig.~\ref{fig:DFSZTherm}.   
Then, imposing that thermalization occurs before $T_{\rm RM}$ limits the parameter space in the $m_{S}$ vs.~$f_{a}$ plane.  
For $n=2$, the interactions rates with Higgs/Higgsino are substantially enhanced for $T>T_S$ as illustrated in the right panel of Fig.~\ref{fig:DFSZTherm}. This offers an opportunity for thermalization at high temperatures. Given that the $\mu$ term scales with temperature as $\mu(T) \propto S^n \propto T^{3n/2}$ in Eq.~(\ref{eq:DFSZDef}), the highest possible thermalization temperature is when $\mu$ is as high as the temperature, $\mu(T_\mu) \equiv T_\mu$. Above this critical temperature $T_\mu$, the Higgs and Higgsino are too heavy to be in thermal equilibrium, so cannot effectively thermalize the saxion. The Higgs and Higgsino interaction rates are parametrically the same at $T_\mu$, differing only by prefactors. Equating the Hubble rate to the interaction rate with Higgs fields at $T_\mu$ yields an upper bound on $f_a$
\begin{equation}
\label{eq:n2Therm}
    f_a \lesssim 2 \times 10^9 \GeV  
    \left(\frac{6}{N_{\rm DW}}\right)
    \left( \frac{\mu}{2 \TeV} \right)^{\frac{1}{2}}
    \left( \frac{g_{\rm MSSM}}{g_*(T_\mu)} \right)^\frac{1}{4}.   
\end{equation}
The requirement that thermalization occurs early enough (either at low temperatures, but before $T_{\rm RM}$, or at $T_\mu$) is non-trivial. We will show its impact on the parameter space below.

\subsection{Conversion of the charge and the generation of baryon number}
The PQ charge associated with the rotation, see Eq.~(\ref{eq:chargedensity}), can be transferred to baryon or lepton number.  In minimal axiogenesis~\cite{Co:2019wyp}, this charge transfer is accomplished via a combination of sphaleron processes.  QCD sphaleron processes convert the PQ charge to chiral quark number.  This is subsequently converted to baryon number via electroweak sphaleron processes.  However, as discussed in the introduction, this charge transfer is found to be not efficient enough once the dark matter abundance constraint on kinetic misalignment is taken into account. An additional transfer mechanism is necessary.  In the present work, this transfer proceeds via the interaction between the right-handed neutrino $N$ and the left-handed leptons $\ell$ and the up-type Higgs field $H_u$
\begin{equation}
\label{eq:Yukawa}
    W_{\nu}= y_N  \ell N H_{u} + \frac{m_N}{2} N^2,
\end{equation}
which can convert the chiral asymmetry of MSSM particles into their $B-L$ asymmetry. Electroweak sphalerons then reprocess the resultant $B-L$ asymmetry to give a final baryon asymmetry $B= \frac{28}{79} (B-L) $ \cite{Harvey:1990qw}.

Right-handed neutrinos can interact with the bath via either scattering or (inverse) decays.  
The scattering rate of right-handed neutrinos with the bath may be estimated by
\begin{equation}
\label{eq:scatteringRate}
   \Gamma_N \sim \frac{g_2^2 y_N^2 T}{8\pi} \sim \frac{\alpha_2 m_\nu m_N T}{2 v^2}.
\end{equation}
Such a scattering process reaches thermal equilibrium before the electroweak phase transition, $\Gamma_N > 3 H$ at $T_{\rm EW}$, for right-handed neutrinos heavier than
\begin{equation}
\label{eq:mN_threshold}
   m_N \gtrsim 20 \GeV 
   \left( \frac{g_*(T_{\rm EW})}{g_{\rm SM}} \right)^{ \scalebox{1.01}{$\frac{1}{2}$} }
   \left( \frac{0.05 \eV}{m_\nu} \right) .
\end{equation}
Similarly, (inverse) decays, $N \leftrightarrow h\nu, Z\nu, \ell^\pm W^\mp$, can be relevant with a total rate of~\cite{Bandyopadhyay:2017bgh}
\begin{equation}
\Gamma_{N,{\rm dec}} = \frac{y_N^2 m_N}{8 \pi} .
\end{equation}

If the right-handed neutrino mass is above the electroweak scale, the interactions of right-handed neutrinos will attain equilibrium in the early universe except when the reheat temperature $T_R$ is low as discussed below. 
The combination of these interactions, along with the Majorana mass term $m_{N}$, effective for $T \sim m_{N}$, provide for $B-L$ violation.

The dynamics of the generation of the baryon asymmetry depend on whether interactions of the right-handed neutrinos achieve equilibrium in the early universe and hence on the mass scale of the right-handed neutrinos. In Sec.~\ref{sec:FreezeOut}, we explore the case where the right-handed neutrino mass is above the weak scale.

In this case, by the electroweak phase transition, the abundance of $N$ becomes negligible. 
Still, a non-zero $B-L$ of MSSM particles generated by the combination of the Majorana mass and the neutrino Yukawa interactions at high temperatures remains.
In Sec.~\ref{sec:DL}, we consider a lighter $N$, for which the Yukawa interaction remains out of equilibrium. The $B-L$ violation by $m_N$ is also ineffective, but the MSSM sector obtains $B-L$ opposite to that of $N$.  The compensating asymmetry stored in $N$ ultimately is of no relevance: when the $N$'s eventually decay, it does not alter the baryon asymmetry. This is because $N$ is a Majorana fermion and the asymmetry in $N$ disappears upon decay, and the decay  anyway occurs after the electroweak phase transition for values of $(y_N,m_N)$ such that the observed neutrino mass is explained.

\section{Lepto-axiogenesis with $m_N$ above the weak scale}
\label{sec:FreezeOut}

In this section, we concentrate on right-handed neutrinos $N$ with masses above the weak scale.  We first discuss the generation of the baryon asymmetry.  Our main result will be finding values of the saxion mass $m_{S}$, axion decay constant $f_{a}$, and right-handed neutrino mass $m_{N}$  that can reproduce the observed asymmetry. We then discuss some complications for the DFSZ axion case.  The consequences of the coupling between $P$ and $H_{u} H_{d}$ can be subtle in the presence of a $P$ field value that changes in time. We then turn to cosmological constraints, many of which arise from the presence of superparters.  Other cosmological constraints that might arise from a possible  parametric resonance of the saxion field and are potentially more model-dependent are postponed to Sec.~\ref{sec:DWandPR}.

\subsection{Production of the asymmetry}
\label{sec:FO_Asymmetry}

When the right-handed neutrinos have masses above the weak scale, their Yukawa interactions with the bath reach equilibrium, see Eq.~\eqref{eq:mN_threshold}.
(An exception is the case with a low reheat temperature, to be discussed in Sec.~\ref{sec:lowTR}.)
When equilibrium is reached, the $B-L$ asymmetry is maintained at the equilibrium value in the presence of the axion rotation, 
\begin{equation}
\label{eq:YBminuL_highTRmN}
    Y_{B-L} = \frac{n_{B-L}}{s} = \frac{c_{B-L} \dot\theta T^2}{s}  ,
\end{equation}
until the Yukawa interactions of $N$ or the electroweak sphaleron processes go out of equilibrium, whichever occurs first. Then $Y_{B-L}$ freezes out and, if entropy is conserved, will remain approximately constant. The above expression should therefore be evaluated at a freeze-out temperature $T_{\rm FO} = \max(x_{\rm FO}^{-1} m_N, T_{\rm EW})$. We find $x_{\rm FO}$ by setting the inverse decay rate (supplemented by a Boltzmann suppression) equal to the Hubble expansion rate, i.e., $\Gamma_{N, {\rm dec}} \times e^{-m_N/T}= 3 H$.  For a radiation-dominated epoch we find $x_{\rm FO} \simeq 6$.   In Eq.~(\ref{eq:YBminuL_highTRmN}), $c_{B-L}$ is a numerical prefactor found by equating the chemical potentials of particles participating in processes that are in thermal equilibrium; for details, see the Appendix of Ref.~\cite{Barnes:2022ren}.%
\footnote{Because the present model has right-handed neutrinos in the bath, the equations are modified with respect to the high-scale lepto-axiogenesis case \cite{Barnes:2022ren}. We remove the equations that set $B/3 - L_i = 0$, and replace them with equations corresponding to the presence of the neutrino Yukawa couplings that set $L_i + H_u = 0$.  Note that the chemical potential for $N$ vanishes for $T \lesssim m_N$, as the scattering and decay involving the Majorana mass term are efficient.}
Assuming all interactions in equilibrium (including the electron Yukawa coupling, applicable when $T \lesssim 10^6 (\tan \beta /3)^2 \GeV$),  we find $c_{B-L} = (15 c_g + 28 c_W)/66$ for the KSVZ model and $c_{B-L} = 185n/(66 N_{\rm DW}) = 185/198$ for the DFSZ model with $c_g$ and $c_W$ denoting the strong and weak anomaly coefficients of the PQ symmetry.  Here, $\tan \beta$ is the ratio of the Higgs vacuum expectation values $\tan \beta \equiv v_{u} / v_{d}$.  For this case where equilibrium is reached, the precise value of the neutrino mass is barely relevant. It enters only logarithmically via the freeze-out  temperature.  For concreteness, we have taken $m_{\nu} = 0.05$~eV.  
 
When evaluating Eq.~(\ref{eq:YBminuL_highTRmN}), it is important to recall that $\dot\theta$ is a function of temperature.  While $\dot \theta$ is initially fixed to $N_{\rm DW} m_S$,  it redshifts once the saxion settles at its minimum.  So, when $x_{\rm FO}^{-1} m_N > T_{\rm EW}$, there exist two values of $m_N$ that reproduce the observed baryon asymmetry, depending on whether the freeze-out of $N$ occurs at a temperature larger or smaller than $T_{S}$. In the latter case, a larger value of $T$ (i.e.~$x_{\rm FO}^{-1} m_{N}$) in Eq.~\eqref{eq:YBminuL_highTRmN} can compensate for the smaller, redshifted value of $\dot{\theta}$. 
  
\begin{figure}
\includegraphics[width=0.495\linewidth]{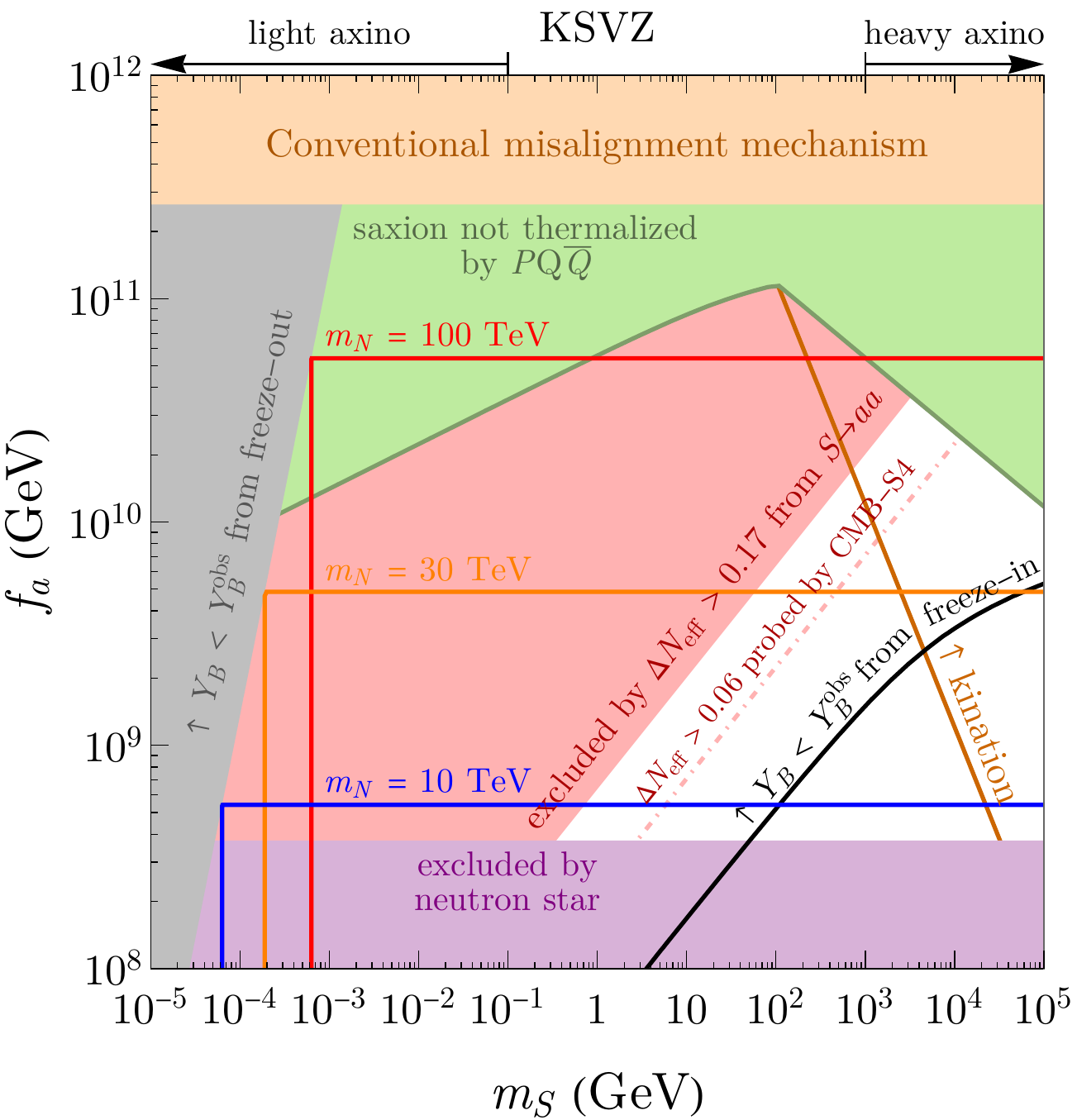} 
\caption{Parameter space for the KSVZ axion for $N_{\rm DW}= 1$ and $c_{B-L} = (15 c_g + 28 c_W)/66$ with $c_g = c_W = 1$ assuming a high $T_R$. Solid contours labeled with $m_N$ show the required parameter values to reproduce the observed baryon asymmetry via lepto-axiogenesis, in the case where the neutrino Yukawa coupling reaches equilibrium, as well as the dark matter abundance via the KMM. Baryon asymmetry is underproduced in the gray regions. The black curve shows the minimum $m_S$ for the out-of-equilibrium case, which is derived in Sec.~\ref{sec:DLProd} using $m_N = 11 \GeV$. The green region shows thermalization constraints discussed in Sec.~\ref{subsec:dynamics}. The red region (red dot-dashed line) corresponds to an excessive (detectable) contribution to dark radiation from saxion decays to axions; see Eq.~(\ref{eqn:NeffFromSaxion}).  This bound is generically present in the KSVZ model, but may be absent in the DFSZ model where more rapid decays via the coupling to the Higgs boson are present; see Sec.~\ref{sec:CRsaxion}. For $m_{S} <$ TeV, a consistent cosmological history for the axino may impose additional model building requirements; see Secs.~\ref{sec:CRaxino} and \ref{sec:lightaxino}. To the right of the orange line, a period of kination occurs prior to the saxion field reaching its minimum.}
\label{fig:KSVZ}
\end{figure}

\begin{figure}
\includegraphics[width=0.495\linewidth]{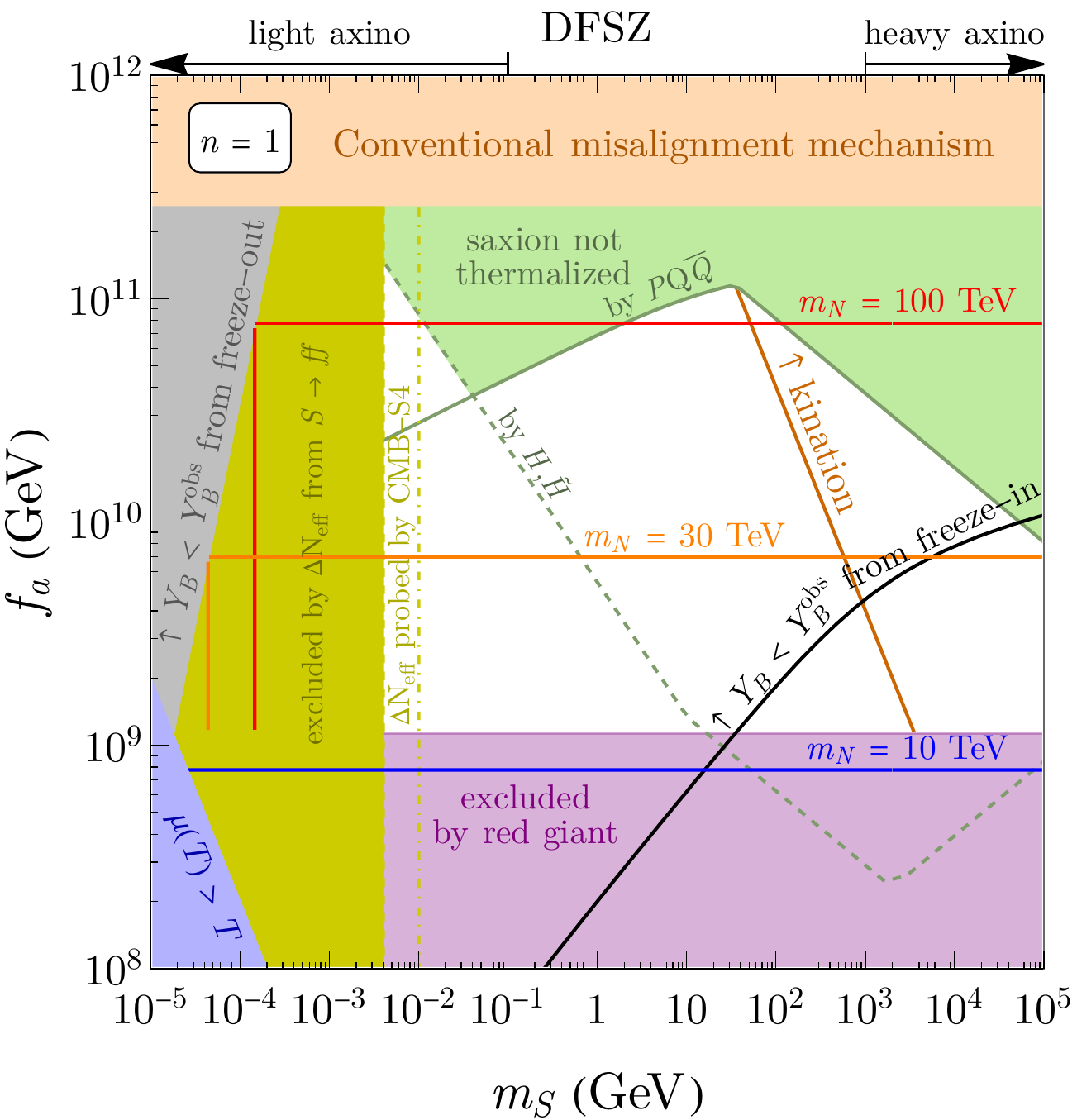} 
\includegraphics[width=0.495\linewidth]{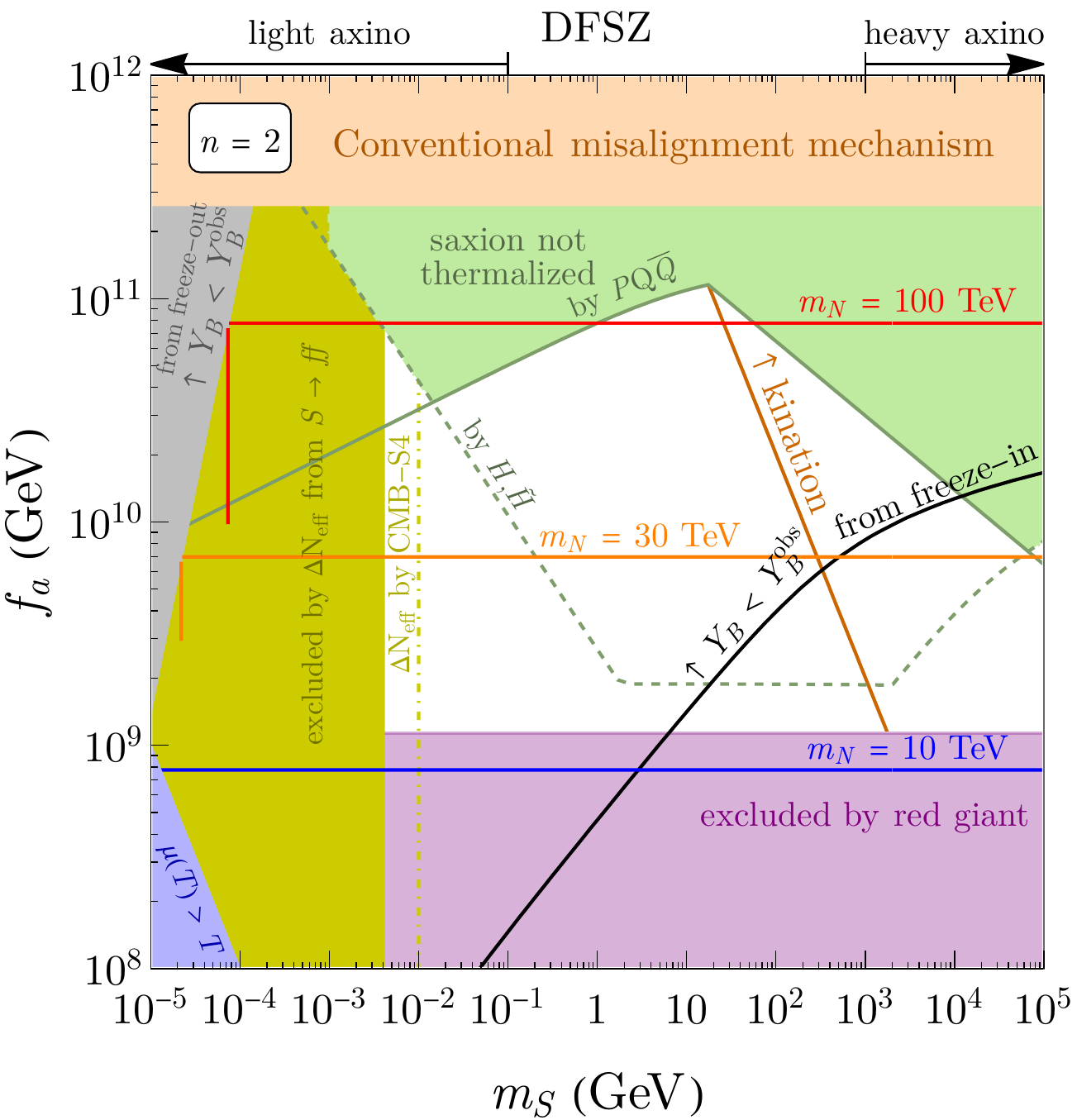} 
\caption{Parameter space for the DFSZ axion,   $W \sim P^{n} H_{u} H_{d}$, with $\mu = \max \left( m_S, 2 \TeV \right)$, $N_{\rm DW}= 3n$, and $c_{B-L} = 185n/(66 N_{\rm DW})$ assuming a high $T_{R}$.  $n =$ 1  ($n=$ 2) for the left (right) panel. When the neutrino Yukawa coupling reaches equilibrium, colored $m_N$ contours reproduce the dark matter abundance via the KMM and the observed baryon asymmetry via lepto-axiogenesis; baryon asymmetry is underproduced in the gray regions. Black curves show the minimum $m_S$ for the out-of-equilibrium case, derived in Sec.~\ref{sec:DLProd} using $m_N = 11 \GeV$. Green regions/lines show thermalization constraints, see Sec.~\ref{subsec:dynamics}. Shaded green regions: the saxion is not thermalized prior to its energy domination even when additional matter $Q$ with favorable couplings to $P$ are present. Above the dashed green lines: saxion does not thermalize via the couplings to the Higgs fields. The yellow region is excluded by {\it Planck}; it gives a too-negative contribution to $\Delta N_{\rm eff}$ from saxion decays to Standard Model fermions~\cite{Ibe:2021fed}; the yellow dot-dashed line shows the future sensitivity of CMB-S4, see Sec.~\ref{sec:CRsaxion}. For $m_{S} <$ TeV, a consistent cosmological history for the axino may impose additional model building requirements; see Secs.~\ref{sec:CRaxino} and \ref{sec:lightaxino}.  To the right of the orange line, a period of kination occurs prior to the saxion field reaching its minimum.}
\label{fig:DFSZ}
\end{figure}

The final baryon asymmetry $Y_B = \frac{28}{79} Y_{B-L}$ is given by
\begin{align}
\label{eq:YBminusL_highTR}
    Y_B & 
    \simeq 9 \times 10^{-11} 
    c_{B-L}
    \begin{dcases}
    \left( \frac{N_{\rm DW} m_S}{\rm MeV} \right)
    \left( \frac{40 \TeV}{x_{\rm FO}^{-1} m_N} \right)
    \left( \frac{g_{\rm MSSM}}{g_*(x_{\rm FO}^{-1} m_N)} \right) 
    & {\rm for} \,\, x_{\rm FO}^{-1} m_N > T_S \\
    \left( \frac{x_{\rm FO}^{-1} m_N}{6 \TeV} \right)^2
    \left( \frac{10^{10} \GeV}{f_a} \right) 
    & {\rm for} \,\, x_{\rm FO}^{-1} m_N < T_S. \\
    \end{dcases}
\end{align}
This is a key result of this work, and allows us to make predictions for the parameters $m_{N}$, $m_{S}$, and $f_{a}$.
Contours of required $m_N$ in the $m_{S}$ vs.~$f_{a}$ plane are shown in Figs.~\ref{fig:KSVZ} and~\ref{fig:DFSZ} that correspond to the KSVZ and DFSZ cases, respectively.  There the vertical/horizontal component of each contour corresponds to the first/second case in Eq.~\eqref{eq:YBminusL_highTR}, respectively. In the DFSZ case, the vertical curves are truncated at low $f_a$ because the assumed particle content in the bath is no longer valid.  We will elaborate on this point in Sec.~\ref{sec:FO_Asymmetry_DFSZ}.

In Fig.~\ref{fig:KSVZ}, the purple region is excluded by constraints on the axion-nucleon coupling that arise from considerations of the neutron star~\cite{Iwamoto:1984ir, Iwamoto:1992jp, Page:2010aw, Shternin:2010qi, Leinson:2014cja, Leinson:2014ioa, Sedrakian:2015krq, Hamaguchi:2018oqw, Leinson:2021ety, Buschmann:2021juv} and SN 1987A cooling~\cite{Ellis:1987pk,Raffelt:1987yt,Turner:1987by,Mayle:1987as,Raffelt:2006cw,Chang:2018rso,Carenza:2019pxu}.
Similarly, the purple region in Fig.~\ref{fig:DFSZ} is excluded by the red giant brightness observations that limit the axion-electron coupling~\cite{Capozzi:2020cbu,Straniero:2020iyi}. The bound on $f_{a}$ is proportional to $\sin^2{\beta}$, which here we set equal to 1. 
We see $x_{\rm FO}^{-1} m_N$ is larger than $T_{\rm EW}$ in the viable parameter space in both figures, so Eq.~(\ref{eq:YBminuL_highTRmN}) is to be evaluated at $T = x_{\rm FO}^{-1} m_N$.
The requirement that $x_{\rm FO}^{-1} m_N> T_{\rm EW}$ can also be simply understood as follows.  When $x_{\rm FO}^{-1} m_N < T_{\rm EW}$, the computation of $Y_{B}$ reduces to that of minimal axiogenesis (with only a minor difference between the $c_B$ of minimal axiogenesis and the $c_{B-L}$ here), and minimal axiogenesis underproduces the baryon asymmetry once astrophysical constraints on $f_a$ are taken into account~\cite{Co:2019wyp}. 

In the gray regions the baryon asymmetry is necessarily underproduced.  There, even after maximizing Eq.~(\ref{eq:YBminuL_highTRmN}) by evaluating it at $T_{\rm FO} = T_S$, the baryon asymmetry is still less than the observed value $Y_B^{\rm obs} = 8.7 \times 10^{-11}$ from {\it Planck}~\cite{Planck:2018vyg}.
To the right of the orange lines labeled kination in Figs.~\ref{fig:KSVZ} and~\ref{fig:DFSZ}, there is an epoch where the energy of rotation comes to dominate, and a period of kination results, see discussions around Eq.~(\ref{eq:TRM}).  Note that if the axino is required to be heavier than the MSSM superpartners then $m_{S}$ is often required to be larger than a TeV,  see discussions in Sec.~\ref{sec:CRaxino}. In this case this kination era is obtained over much of the allowed parameter space, especially in for a DFSZ axion.

Considerations of thermalization also limit the allowed parameter space, see Sec.~\ref{sec:Therm}.  In the regions shaded green in Figs.~\ref{fig:KSVZ} and \ref{fig:DFSZ}, the saxion is not successfully thermalized before $T_{\rm RM}$, even in the presence of a field $Q$ with favorable couplings to the saxion.  The shaded regions above the positively-sloped segments of the solid green lines correspond to the discussion above Eq.~(\ref{eq:thermRadiation}) when the thermalization happens in a radiation dominated era before the saxion has a chance to dominate. As discussed around Eq.~(\ref{eq:thermTRM}), thermalization occurs before the saxion dominates below the negatively sloped segments, while thermalization after the saxion domination fixes the parameters to lie along the negatively sloped segments. Points above this line do not reproduce the correct axion dark matter abundance.  We emphasize that the presence of a period of saxion domination does not modify the baryon asymmetry evaluation as long as the freeze-out temperature $x_{\rm FO}^{-1} m_N$ is lower than the temperature $T_{\rm th}$ where thermalization occurs. This is true for the horizontal components of the colored contours, i.e., the first case in Eq.~\eqref{eq:YBminusL_highTR}.   The vertical components (the second case) would require a new evaluation along the negatively-sloped segment to account for entropy production from the saxion.  However, for points well below this contour the entropy would again be negligible. This will not be pursued further here.

We now discuss the range of applicability of Eq.~\eqref{eq:YBminusL_highTR}. We have so far implicitly assumed that $m_N$ is sufficiently light that it is in the bath when the PQ field begins oscillation/rotation, i.e., $x_{\rm FO}^{-1} m_{N} < T_{\rm osc} \sim \sqrt{m_S M_{\rm Pl}}$.
By using $x_{\rm FO}^{-1} m_N \le T_{\rm osc}$ and the first line of Eq.~(\ref{eq:YBminusL_highTR}), we can determine that this assumption is valid for
\begin{equation}
\label{eq:FO_after_Tosc}
    m_S \lesssim \frac{100 \GeV}{(c_{B-L} N_{\rm DW})^2},
\end{equation}
and $m_N \lesssim 2 \times 10^{10} \GeV/(c_{B-L} N_{\rm DW})$. If these inequalities are not satisfied, $B-L$ is no longer dominantly produced after rotations have begun in earnest.  Instead, the bulk of production occurs at an earlier time when $N$ is in the bath but when $T > T_{\rm osc}$. 
Because the mass of $P$ is smaller than $H$, $P$ is (slowly) moving.  During this slow-roll phase, the angular velocity of $P$ induced by Eq.~\eqref{eq:VAterm} is still as large as $A\sim m_S$. This means, even though the dynamics differs, Eq.~\eqref{eq:YBminusL_highTR} applies to even higher masses than suggested by Eq.~\eqref{eq:FO_after_Tosc}.
For sufficiently large $m_S$, however, the production again dominantly occurs while $P$ rapidly rotates, for which the neutrino mass term should be treated as the effective dimension-five Majorana mass $(\ell H)^2$, and the scenario reduces to high-scale lepto-axiogenesis at $m_S \simeq \mathcal{O}(10) \TeV$~\cite{Co:2020jtv, Barnes:2022ren}. For even larger $m_S$, the baryon asymmetry is overproduced.

It is not only $N$ that can potentially be absent from the bath; in the DFSZ model, the Higgs fields may be out of equilibrium if the $\mu$ term is larger than temperature in the early times. This possibility is discussed below in Sec.~\ref{sec:FO_Asymmetry_DFSZ}.

\subsection{Complications for the DFSZ Axion}
\label{sec:FO_Asymmetry_DFSZ}

DFSZ axions are distinguished by the presence of the $P^n H_{u} H_{d}$ coupling in Eq.~(\ref{eq:DFSZDef}).
 For high temperatures (above $T_{S}$),  $P$ has a larger field value so that the $\mu(T)$ term may be enhanced.  Because we have been considering asymmetry generation at $T\approx x_{\rm FO}^{-1} m_{N}$, for large values of $m_{N}$ this may be an important effect.  The concern is that the effective $\mu$ term may be larger than the freeze-out temperature, i.e.,  $\mu(T) > x^{-1}_{\rm FO} m_{N}$.  Then the Higgs fields will not be in the bath at $x^{-1}_{\rm FO} m_{N}$, and  generation of the asymmetry would proceed via higher-dimensional operators suppressed by this large $\mu(T)$. This complication is avoided~for 
 \begin{align}
\label{eq:fa_trunc1}
 f_a & \gtrsim  10^9 \GeV 
 \left( \frac{c_{B-L}}{185/198} \right) 
 \left( \frac{\mu}{2 \TeV} \right)^2 
 & \text{for} & \hspace{1cm} n = 1,  \\
 f_a & \gtrsim  10^{12} \GeV 
\left( \frac{c_{B-L}}{185/198} \right)^2 
 \left( \frac{\mu}{2 \TeV} \right)
 \left( \frac{m_S}{10 \MeV} \right) 
 \left( \frac{N_{\rm DW}}{6} \right)
 & \text{for} & \hspace{1cm} n = 2.
 \end{align}
 Below these values, the calculation of the baryon asymmetry changes. Recall, this is relevant for the case of freeze-out before $T_S$ (first case in Eq.~\eqref{eq:YBminusL_highTR}). We have therefore truncated the vertical lines in the left and right panel of Fig.~\ref{fig:DFSZ} accordingly. 
  For $n=1$ and $\mu \lesssim 2$ TeV, this results in no additional bounds beyond those from red giant brightness observations. We note that, as discussed below Eq.~(\ref{eq:FO_after_Tosc}), $m_S > \mathcal{O}(10) \TeV$ reduces to high-scale lepto-axiogenesis. For $n=2$, the truncation is such that the vertical branch never reaches values of $m_S$ and $f_a$ that evade other constraints, even for  $m_N$ larger than the benchmark values used in the figure.  So for $n=2$, only the second expression in Eq.~(\ref{eq:YBminusL_highTR}) (i.e., horizontal colored lines) is viable.
 The $\mu$ term can also be larger than $T$ throughout the entire cosmological evolution. This applies if $\mu > T_S$, which leads to the blue-shaded region in Fig.~\ref{fig:DFSZ}, and the second case in Eq.~(\ref{eq:YBminusL_highTR}), namely freeze-out after $T_S$, is invalid in this region.

We now turn to thermalization constraints in the DFSZ model where thermalization could occur even in the absence of new PQ-charged matter.  If the minimal DFSZ matter content is present, thermalization occurs via interactions with the Higgs fields, see Sec.~\ref{sec:thermHiggs}, and the region in Fig.~\ref{fig:DFSZ} that does not thermalize is instead bounded by the green dashed lines.  Along the shallower negatively-sloped and positively-sloped segments in the left panel of Fig.~\ref{fig:DFSZ}, scattering with Higgsinos is the dominant thermalization channel. The slope changes at $m_S=2$ TeV because we set $\mu = \max(m_S, 2 \TeV)$. Along the steeper negatively-sloped segments, Higgs scattering dominates. This also sets the negatively-sloped segment of the green dashed line in the right panel, where $n=2$ is assumed.  As discussed near Eq.~(\ref{eq:n2Therm}), substantial enhancement of the $P$-Higgs interactions due to the field-dependent $\mu$ term for $T> T_{S}$ modifies the story for $n=2$.  This corresponds to the horizontal and positively-sloped segments of the green dashed curve in the right panel; we again take $\mu = \max(m_S, 2 \TeV)$.

\begin{figure}
\includegraphics[width=0.495\linewidth]{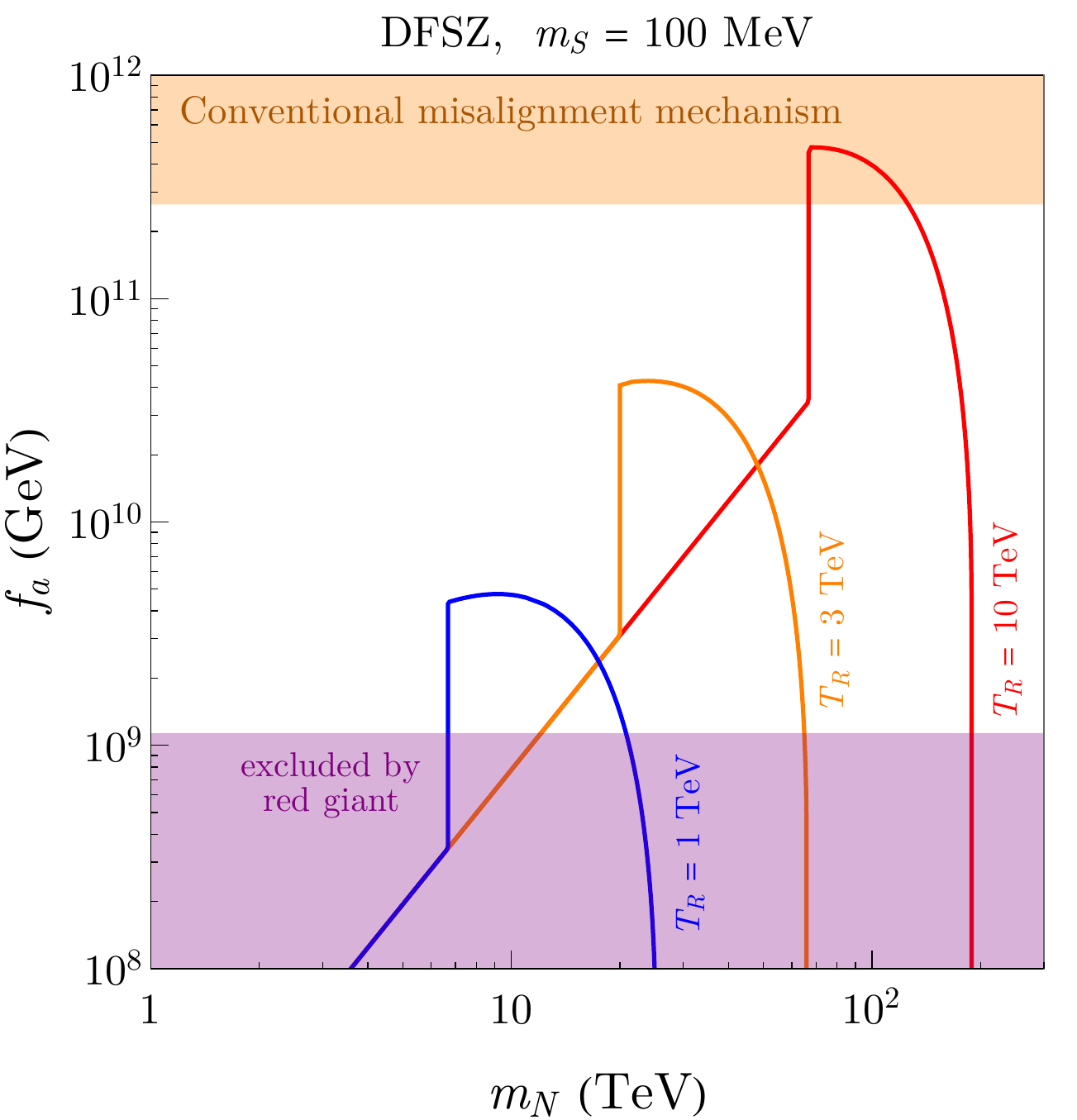} 
\includegraphics[width=0.495\linewidth]{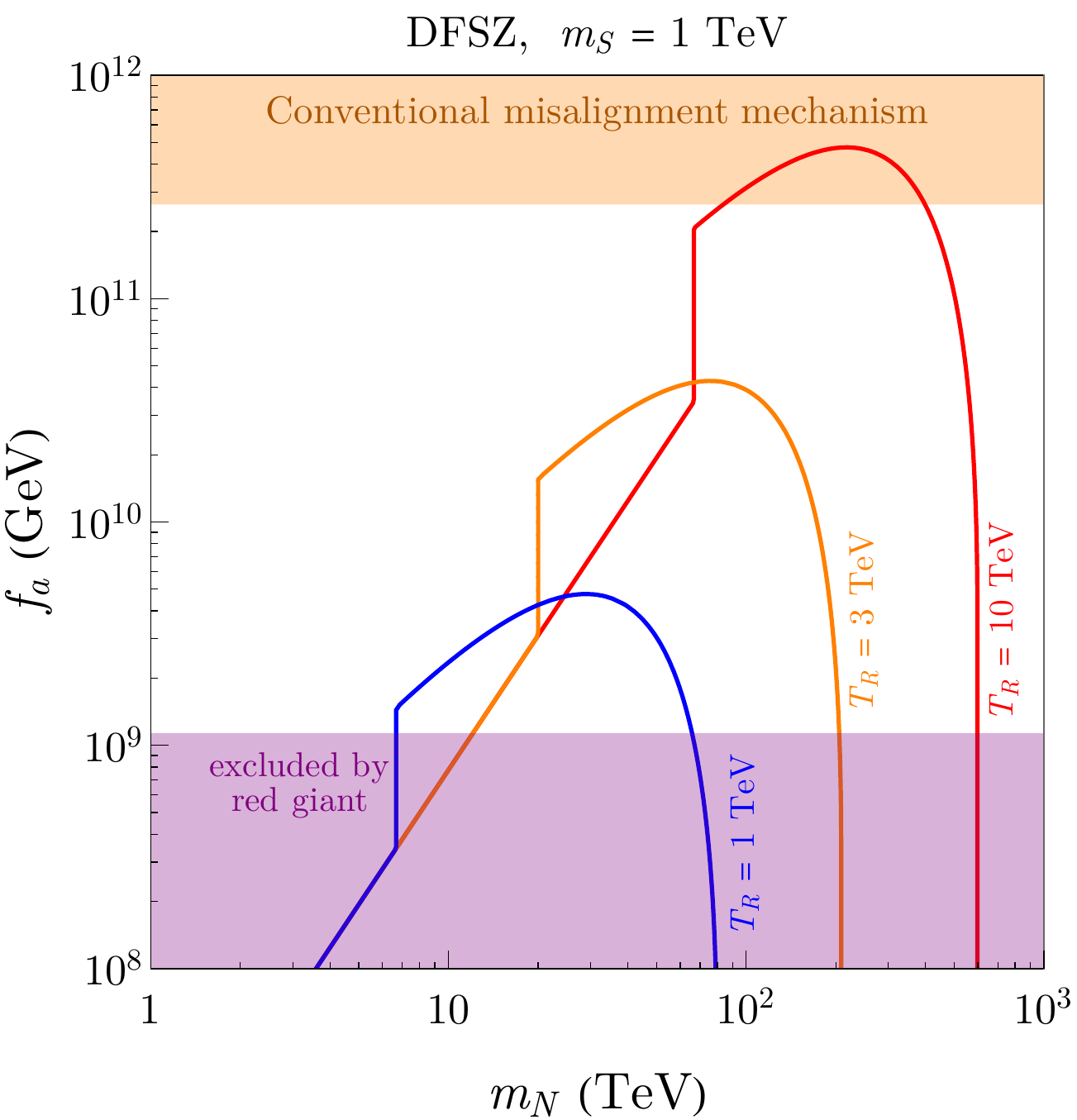} 
\caption{Parameter space for the DFSZ axions in the case where the reheat temperature is low. The left (right) panel assumes $m_S = 100 \MeV$ ($m_S = 1 \TeV$). Colored $T_{R}$ contours reproduce the dark matter abundance via the KMM and the observed baryon asymmetry via lepto-axiogenesis for the labeled values of $T_R$.}
\label{fig:DFSZ_TR}
\end{figure}

\subsection{Generation of the baryon asymmetry for low reheat temperatures}
\label{sec:lowTR}
Thus far, we have implicitly assumed that the reheat temperature following inflation $T_R$ is higher than $T_{\rm FO} = x_{\rm FO}^{-1} m_N$. In some supersymmetric models, $T_R$ has a stringent upper bound from the gravitino problem.  This will be discussed in more detail in Sec.~\ref{sec:cosmologicalrelics}.  A low reheat temperature could even mean $T_R < T_{\rm FO}$, in which case the production of the asymmetry still stops at $T = T_{\rm FO}$, but this now occurs \emph{during} the reheating era.  
The production of entropy during reheating  dilutes the asymmetry. The redshift-invariant quantity from $T=T_{\rm FO}$ to the end of reheating is not $n_{B-L}/s$ but rather $n_{B-L}/\rho_{\rm inf}$ with $\rho_{\rm inf}$ the inflaton energy density. During this period the Hubble rate is enhanced by the inflaton energy density.  This typically results in right-handed neutrino interactions being out of equilibrium, $\Gamma_N \ll H$, and thus the $n_{B-L}$ production is in the ``freeze-in" regime. We find that the freeze-in picture applies shortly after $T_R$ falls below $T_{\rm FO}$, so a freeze-out during matter-domination  is rarely the case. For freeze-in, the production of $B-L$ asymmetry per Hubble time is suppressed by the ratio $\Gamma_N / H$. The production tends to be dominated at low temperatures, where the relevant rate is $\Gamma_{N,{\rm dec}}$ accompanied by a Boltzmann suppression factor $e^{-m_N/T}$ that applies  for $T \lesssim m_N$.
Allowing for the freeze-in possibility and accounting for dilution, the final baryon asymmetry is given by
\begin{equation}
\label{eq:YBminusL_lowTR}
    Y_B = \frac{28}{79} \left. \frac{n_{B-L}}{\rho_{\rm inf}} \frac{\Gamma_{N,{\rm dec}}}{H} e^{-m_N/T} \right|_{T = T_{\rm peak}} \hspace{-0.5cm} \times \left. \frac{\rho_{\rm inf}}{s} \right|_{T=T_R} 
\hspace{1cm} {\rm for}~~~ T_{R} < T_{\rm peak}.
\end{equation}
Here $T_{\rm peak}$ refers to the temperature at which $Y_B$ production peaks. The temperature dependence of $n_{B-L}$ is given by $\dot\theta T^2$ with $\dot\theta = N_{\rm DW} m_S \times \min(1,(T/T_S)^8)$ with $T_S$ given in Eq.~(\ref{eq:TS_KMM_TR}), while $\rho_{\rm inf}$ and $H \propto \sqrt{\rho_{\rm inf}}$ depend on temperature as $T^8$ and $T^4$, respectively. We have used the temperature dependence of the scale factor $R^{-3} \propto T^8$, assuming a perturbative decay of the inflaton. Using these temperature scalings, we find three possibilities for $T_{\rm peak}$ during the inflationary era. First, $N$ may fall out of equilibrium well before the rotation settles to the minimum at $T=T_S$. Then $Y_{B-L}$ production is IR-dominated until $T_{\rm peak} = x_{\rm FI}^{-1} m_N > T_S > T_R$. In this case, we obtain $x_{\rm FI} \simeq 10$ from maximizing $Y_B$ in Eq.~(\ref{eq:YBminusL_lowTR}). A second possibility is that $N$ may fall out of equilibrium after $T_S$, and the $Y_{B-L}$ production is IR-dominated until $T_{\rm peak} = x_{\rm FI}^{-1} m_N > T_R$, and $x_{\rm FI} \simeq 2$ when $T_S \gg T_{\rm peak}$. Third, $N$ may fall out of equilibrium right around $T_S$ ($m_N \gtrsim T_S \gtrsim m_N/10$, so the previous two cases do not apply); in this case production simply peaks at $T_{\rm peak} = T_S > T_R$.

In Fig.~\ref{fig:DFSZ_TR}, contours that reproduce the baryon asymmetry are shown in the low $T_R$ scenario for the DFSZ axion. These panels fix $m_\nu = 0.05 \eV$ while varying $T_R$ in the different colored contours. The left and right panels are for $m_S=100 \MeV$ and $m_S = 1 \TeV$, respectively. We now describe the physics of the colored contours, starting at the lower left and moving to the right. The straight sloped segments correspond to the horizontal segments of Fig.~\ref{fig:DFSZ}. This is the case where the dominant production occurs at $T_{\rm FO} = x_{\rm FO}^{-1} m_N < T_S$ according to the second case of Eq.~(\ref{eq:YBminusL_highTR}); the $f_a$ dependence arises through the $\dot\theta(T)$ dependence on $T_S$. 
The left vertical segments correspond to production transitioning from the freeze-out to the freeze-in case, occurring at $T_R \simeq x_{\rm FO}^{-1} m_N$.  In this case, the temperature where the $B-L$ asymmetry is dominantly produced is very sensitive to changes in $m_N$, so the required $f_a$ changes rapidly. The curved segments correspond to the freeze-in case; there the $\Gamma_{N,{\rm dec}} e^{-m_N/T} / H$ suppression is relevant. Along the positively sloped curved segment (more prominent in the right panel), we have $T_{\rm peak} = x_{\rm FI}^{-1} m_N$, i.e., the {\it second} possibility discussed below Eq.~\eqref{eq:YBminusL_lowTR}, while the remainder of the curved segment has $T_{\rm peak} = T_S$, i.e., the {\it third} possibility. Lastly, the right vertical segments below the curved parts correspond to freeze-in production at $T = x_{\rm FI}^{-1} m_{N} > T_S$, i.e., the {\it first} possibility discussed below Eq.~\eqref{eq:YBminusL_lowTR}, where $\dot\theta = N_{\rm DW} m_S$ so that $T_S$ and hence $f_a$ become irrelevant. The transition from the curved segment to this vertical segment therefore occurs when $x_{\rm FI}^{-1} m_{N} = T_{S}$. As a result, as we vary from $m_S = 100 \MeV$ to $m_S = 1 \TeV$ between the two panels, we see the rightmost vertical segments shift to the right according to $m_N \propto T_S \propto m_S^{1/8}$; see Eq.~(\ref{eq:TS_KMM_TR}) for $T_S$ during reheating. The sloped and left vertical segments as well as the maximum $f_{a}$ reached for each contour remain unchanged. 

We do not display an analogous low $T_{R}$ figure for the KSVZ axion.  The reason is as follows: the region that most strongly favors a low $T_{R}$ scenario is low $m_S$, where the bounds coming from the saxion cosmology, see Sec.~\ref{sec:CRsaxion}, are particularly strong for the KSVZ axion.  These are shown in the red region of Fig.~\ref{fig:KSVZ}.  It is, however, possible that a low reheating temperature could apply to the KSVZ case for a TeV scale $m_{S}$.  In this case, the figure looks very similar to the right hand panel of Fig.~\ref{fig:DFSZ_TR}, with the small numerical values primarily due to shifts in the domain wall number or the value of $c_{B-L}$, see Eq.~(\ref{eq:YBminuL_highTRmN}).

A sufficiently low reheat temperature may be inconsistent with the KMM as it tends to reduce the PQ charge density. When the rotation begins with an initial field value $S_i$, the ratio between the PQ charge and the inflaton energy density $\rho_{\rm inf}$ is $m_S S_i^2 / (m_S^2\MPl^2)$, which remains constant until the reheating completes. After the completion of the reheating, the yield of the PQ charge is
\begin{align}
\left| Y_\theta \right| = \left| \frac{n_\theta}{\rho_{\rm inf}} \frac{3T_R}{4} \right|= \frac{9 S_i^2 T_R}{4 m_S\MPl^2} < \frac{9T_R}{4m_S},
\end{align}
where we impose $S_i < \MPl$ in the inequality. Requiring that the dark matter abundance be produced by KMM then puts an upper bound on $f_a$,
\begin{equation}
\label{eq:lowTRfabound}
f_a \lesssim 3 \times 10^{11} \GeV
\left(\frac{T_R}{\rm TeV} \right)
\left(\frac{100 \MeV}{m_S}\right). \hspace{.5 in} \textnormal{(KMM DM,  low $T_{R}$)}
\end{equation}
The bound is strong for large $m_S$ and/or low $T_R$.  So, when considering the low $T_{R}$ scenario, relatively low $m_{S}$ will be of  most interest.  Note that the saxion mass to be used in Eq.~(\ref{eq:lowTRfabound}) is the saxion mass  around $S_i$ rather than that around the minimum of the potential. Thus, the upper bound on the decay constant can be weaker than this using $m_S$ at the minimum if the saxion potential becomes flatter at large field values. Such a case is realized if the saxion potential is generated by dynamics much below the Planck scale. A flat potential, however, will result in the production of Q-balls~\cite{Coleman:1985ki,Kusenko:1997zq,Kusenko:1997si,Kasuya:1999wu} at some stage of the evolution of the rotation. Although the Q-balls may eventually do decay, the cosmological impact of the Q-balls should be carefully examined, which is beyond the scope of this paper.

\subsection{Cosmological relics}
\label{sec:cosmologicalrelics}

The theory has several long-lived or stable particles. They can be produced by the scattering in the thermal bath and may cause cosmological problems. In this subsection, we discuss the saxion, axino, gravitino, right-handed neutrino and sneutrino, as well as the lightest observable supersymmetric particle (LOSP).  When we refer to the LOSP, we mean the lightest MSSM superpartner excluding the gravitino, axino, and right-handed sneutrinos.  Many of the results of these discussions are also summarized in Figs.~\ref{fig:KSVZ} and \ref{fig:DFSZ}.

\subsubsection{Saxion}
\label{sec:CRsaxion}
When the rotation is thermalized, the saxion is also necessarily thermalized and obtains a thermal number density $\sim T^3$. The decays of the saxion are model dependent, but it decays to axions typically with a rate 
\begin{align}
\Gamma_{S\rightarrow a a} = \frac{m_{S}^3}{32 \pi N_{\rm DW}^2 f_a^2} .
\end{align}
If this decay occurs after the saxion dominates the energy density of the bath, it can potentially produce too much dark radiation.%
\footnote{In the two-field model with $m_P \simeq m_{\bar{P}}$ decays to the axion will be suppressed. In this limit, there is an approximate $P \leftrightarrow \bar{P}$ symmetry.  Under this symmetry, $S \rightarrow -S$ and $a\rightarrow -a$, so the $Saa$ vertex is forbidden.  The saxion would dominantly decay to a pair of SM gauge bosons; the dark radiation bound can be avoided. However, inside the red region of Fig.~\ref{fig:KSVZ}, the decay occurs after $T_{\rm RM}$. Entropy is produced via the decays, diluting the baryon asymmetry.} 
Assuming the above decay dominates, as is relevant for the KSVZ axion, the resultant $\Delta N_{\rm eff}$ is
\begin{align}
\Delta N_{\rm eff} \simeq 0.3 \,
N_{\rm DW}
\left( \frac{\rm GeV}{m_S} \right)^{ \scalebox{1.01}{$\frac{1}{2}$} }
\left( \frac{f_a}{10^9 \GeV} \right)
\left( \frac{100}{g_*(T_D)} \right)
\left( \frac{10}{g_*(T_{S\rightarrow aa})} \right)^{ \scalebox{1.01}{$\frac{1}{12}$} }. 
\label{eqn:NeffFromSaxion}
\end{align}
Here we have taken into account the effect of the change of $g_*$ assuming that the saxion first decouples with the thermal bath at the temperature $T_D$ (while relativistic) and later decays to axions at the temperature $T_{S \rightarrow aa}$.
The red region labeled by $\Delta N_{\rm eff} > 0.17$ in Fig.~\ref{fig:KSVZ} corresponds to the region where the hot axions exceed the current bound on dark radiation from the BBN and CMB observations~\cite{Fields:2019pfx}. The region above the red dot-dashed lines can be probed by CMB-S4, whose projected sensitivity is $\Delta N_{\rm eff} \le 0.06$ at 95\% confidence level~\cite{Abazajian:2019eic}. In the DFSZ model, on the other hand, the saxion directly couples to the Higgs bosons,
\begin{align}
\mathcal{L}_{SHH} = \frac{2n}{N_{\rm DW}} \frac{\mu^2}{f_a}|H|^2 S.
\end{align}
This coupling can induce saxion decays to (longitudinal) pairs of gauge bosons and Higgs bosons if kinematically accessible. This coupling also induces saxion--Higgs mixing, which gives decays to Standard Model fermions should the decays to electroweak bosons not be accessible. While there are also scattering processes,  at the temperature $T=m_{S}$ we expect them to be of comparable importance to the decay rates considered here and to not significantly change conclusions. Depending on the value of $\mu$, these decays can be more rapid, inducing saxion decays prior to the epoch when the saxion becomes non-relativistic, thus relaxing the dark radiation constraint.   If we require the decay rate to exceed three times the Hubble rate at $T=m_{S}$, we find
\begin{align}
\label{eq:muFromSaxion}
    \mu > 700 \GeV 
    \left( \frac{N_{\rm DW}}{q_\mu n} \right)^{ \scalebox{1.01}{$\frac{1}{2}$} }
    \left(\frac{g_*}{10} \right)^{ \scalebox{1.01}{$\frac{1}{8}$} }
    \left( \frac{m_{S}}{50 \GeV}\right)^{ \scalebox{1.01}{$\frac{1}{4}$} }   
    \left(\frac{f_a}{ 10^{10} \GeV}\right)^{ \scalebox{1.01}{$\frac{1}{2}$} }
    \left(\frac{3}{N_f}\right)^{ \scalebox{1.01}{$\frac{1}{4}$} }
    \left(\frac{m_{\rm bottom}}{m_f}\right)^{ \scalebox{1.01}{$\frac{1}{2}$} } ,
\end{align}
where $m_f$ is the mass of the fermions the saxion can decay into, and $N_f$ is the fermion multiplicity~\cite{Co:2017orl}. The constraint when $m_S > 2 m_Z$ is even weaker.
So, for large enough $\mu$ the constraint on $\Delta N_{\rm eff}$ is absent in the DFSZ model, and the red region of Fig.~\ref{fig:KSVZ} need not apply.  This explains its absence in Fig.~\ref{fig:DFSZ}.  However, if $m_S \lesssim 4$ MeV, even if $\mu$ satisfies the constraint in Eq.~\eqref{eq:muFromSaxion}, $S$ has a non-negligible thermal abundance when the neutrinos decouple from the bath. The saxion decay subsequently injects entropy to photons, and this leads to a \emph{negative} contribution to $\Delta N_{\rm eff}$~\cite{Ibe:2021fed}. The resulting constraint is shown by the yellow regions in Fig.~\ref{fig:DFSZ}.

 Notably, the minimum values of $m_S$ are set by these constraints from $\Delta N_{\rm eff}$. For the DFSZ model, the yellow region constrains $m_S \gtrsim 4 \MeV$ and thus sets a minimum value of $m_{N} \simeq n \times 3$~PeV  for the case with freeze-out before $T_S$ (vertical components). For the KSVZ model, the red region requires $m_S > 0.3 \GeV$ and accordingly necessitates $m_N > 50$~PeV.
 On the other hand, the case with freeze-out after $T_S$ allows for $m_N = \mathcal{O}(10-100) \TeV$ indicated by the horizontal components of both DFSZ and KSVZ figures.

\subsubsection{Axino}
\label{sec:CRaxino}

Next, we consider cosmological constraints arising from the production of the axino, the fermionic superpartner of the axion. The axino will be thermalized when the rotation is thermalized, since the interactions of the axino are related to those of the saxion via supersymmetry.%
\footnote{In the DFSZ model, the thermalization of the axino may be avoided if the Higgsino mass is much above the weak scale and the thermalization of the rotation occurs after the Higgsino decouples. We find that this is possible only for $m_S < 0.1$ GeV.  Furthermore, the axino is still produced by freeze-in processes and the upper bound on the mass of a stable axino discussed below, $m_{\tilde{a}} < 10$ eV,  is relaxed at the most by a factor of 100.
}
This thermal abundance can have important implications; but they depend upon the mass of the axino. The axino mass typically satisfies $m_{\tilde{a}} < m_S$ \cite{Chun:1992zk}.
 The equality is saturated, for example, when both the saxion and axino masses come from tree-level gravity-mediated effects.

For $m_S >$ TeV, the axino may be heavier than the MSSM particles and can decay into them. In the DFSZ model, the axino decays into a Higgs boson and a Higgsino with a rate
\begin{equation}
\label{eq:DFSZaxinotoHiggs}
\Gamma_{\tilde{a}}^{\rm DFSZ}=\frac{n^2}{16\pi N_{\rm DW}^2} \frac{\mu^2 m_{\tilde{a}}}{f_a^2}.
\end{equation}
If this is sufficiently large, the axino decays before the Higgsino undergoes freeze-out, and these decays do not overproduce MSSM dark matter. We find 
\begin{equation}
\frac{\Gamma_{\tilde{a}}^{\rm DFSZ}}{H_{\rm fo}} =  \left(\frac{7 \times 10^{10} \ GeV}{f_a}\right)^2 \left(\frac{m_{\tilde{a}}}{\rm{TeV}}\right)\frac{n^2}{N_{DW}^2},
\end{equation}
where we have evaulated the Hubble expansion rate at freezeout $H_{\rm fo}$ at a temperature $T_{\rm fo} \simeq \mu/20$.  

For the KSVZ case, the decay of the axino is less rapid.  It instead decays to a $SU(N)$ gauge boson and its associated gaugino with rate 
\begin{equation}
\Gamma_{\tilde{a}}^{\rm KSVZ} = \alpha_{N}^2 \frac{N^2 -1}{128 \pi^3} \frac{m_{\tilde{a}}^3}{f_a^2},
\end{equation}
where we have neglected possible kinematic suppression.  If we consider decays to a wino Lightest Supersymmetric Particle (LSP), we find
\begin{equation}
    \frac{\Gamma_{\tilde{a}}^{\rm KSVZ}}{H_{\rm fo}} =  \left(\frac{8\times 10^{8} \rm{\; GeV}}{f_a}\right)^2 \left(\frac{m_{\tilde{a}}}{3 \rm{\; TeV}}\right)\left(\frac{m_{\tilde{a}}}{m_{\tilde{w}}}\right)^2.
\end{equation}
Alternately, if the gluino is kinematically accessible, we have
\begin{equation}
    \frac{\Gamma_{\tilde{a}}^{\rm KSVZ}}{H_{\rm fo}} =  \left(\frac{4 \times 10^{9} \rm{\; GeV}}{f_a}\right)^2 \left(\frac{m_{\tilde{a}}}{3 \rm{\; TeV}}\right)\left(\frac{m_{\tilde{a}}}{m_{\tilde{g}}}\right)^2.
\end{equation}
In summary, the requirement of sufficiently rapid axino decay for TeV-scale axino can non-trivially constrain the allowed region of $f_{a}$, but the constraint depends both on the choice of the axion model and the details of the superpartner spectrum.  The constraints on the DFSZ model are mild.

For $m_S <$  TeV, the axino will be lighter than the LOSP and the right-handed sneutrino.\footnote{The right-handed sneutrino cannot be the LSP, see Sec.~\ref{sec:CRsneutrino}.} Assuming $R$-parity conservation, the only open decay is to an axion and gravitino.\footnote{We find that RPV decays are slow. The axino would  decay after BBN, even in the presence of RPV.} Unless the gravitino mass is below 10 eV, the abundance of gravitino resulting from these decays is both too large and will result in hot dark matter. It is difficult to obtain the MSSM soft masses at the TeV-scale with a gravitino mass below 10 eV. (See, however, Ref.~\cite{Hook:2015tra}.)   So, an axino mass below a TeV seems challenging to reconcile with cosmological constraints.  We have therefore marked the viable region with $m_{S}>$ TeV as ``heavy axino."

However, the situation may change if the axino is quite light, with mass below 10 eV.  In this case, even a thermalized axino could be consistent with cosmological constraints.   We return to a discussion of this possibility, along with some related model-building challenges in Sec.~\ref{sec:lightaxino}.

\subsubsection{Gravitino}
\label{sec:Gravitino}

\emph{Stable Gravitinos:}
The gravitino can be produced by the scattering from the thermal bath or by decay of supersymmetric particles.
If the gravitino is the LSP, the gravitino becomes dark matter and $T_R$ is bounded~\cite{Kawasaki:2008qe},
\begin{align}
\label{eqn:BBN_RH}
T_R < 2\times 10^8~{\rm GeV}~\left(\frac{m_{3/2}}{100~{\rm GeV}}\right)  \left( \frac{3~{\rm TeV}}{M_3}\right)^2,
\end{align}
where we assume the production is dominated by the scattering with/decay of the gluino. 

Since the gravitino is lighter than the LOSP in this case, there are also potential constraints that arise from the LOSP decay into the gravitino. Such decays must not disturb BBN~\cite{Kawasaki:2008qe}.  One possibility is that the decays simply occur before BBN.  The decay rate to the gravitino goes like $1/F^2 \propto m_{3/2}^{-2}$ with $F$ the scale of supersymmetry breaking. For a TeV scale bino or wino NLSP, this would require a relatively light gravitino, with $m_{3/2} \lesssim 10$~GeV.  Then, precisely because the interactions of the gravitino are relatively unsuppressed, the thermal production of the gravitino is large and the constraint on the reheat temperature $T_{R}$, Eq.~(\ref{eqn:BBN_RH}) is strong.  Indeed, $T_{R}$ may be constrained to be less than $T_{S}$.
The constraints in the case of the stau NLSP are slightly weaker and come from bounds that result from the catalysis of lithium. A TeV scale stau with $m_{3/2} \lesssim$ 30 GeV gravitino avoids these bounds, thus allowing slightly larger $T_{R}$. Similarly, if the sneutrino is the LSP, the decays to the gravitino and neutrino are innocuous.  However, the subdominant 3-body decay that includes a $Z$-boson can still be important and, again, should limit the gravitino to be $\lesssim {\mathcal O (10)}$ GeV.

Moreover, the abundance of the gravitino that results from the decays should not exceed the dark matter density.  If a mild hierarchy between the LOSP and the gravitino exists, with a LOSP of mass roughly a TeV and a gravitino mass of a few 10's of GeV as discussed above, we expect this should be satisfied.  

In summary, a stable gravitino requires a somewhat low reheat temperature, see Eq.~(\ref{eqn:BBN_RH}), as well as a hierarchy between the gravitino and the LOSP $m_{3/2} < {m_{\rm SUSY}}$.

\emph{Unstable Gravitinos:} 
Another possibility is that the gravitino is heavier than the LOSP. In this case, it decays into the MSSM particles. These decays also have the potential to disturb BBN, with a resulting bound on $T_R$ of $10^{5-9}$ GeV for $m_{3/2}=10^{3-4}$ GeV~\cite{Kawasaki:2008qe,Kawasaki:2017bqm}. 

Comparing with Eq.~\eqref{eq:YBminusL_highTR}, we note that $T_R$ may be lower than $m_N$, for which case the computation in Sec.~\ref{sec:lowTR} should be applied, see Fig.~\ref{fig:DFSZ_TR} for results.
On the contrary, to satisfy the high-reheat temperature assumption (as in Figs.~\ref{fig:KSVZ} and \ref{fig:DFSZ}), a mild hierarchy $m_{3/2} \gtrsim $ few TeV $ > m_{S}$ is required for $m_{S} \lesssim$ TeV. For now we have assumed that the axino is heavy, ${\mathcal O}$(TeV).  We will comment on the possibility that the axino is light (which would allow gravitino decays) in Sec.~\ref{sec:lightaxino}.

\subsubsection{Right-handed (s)neutrino}
\label{sec:CRsneutrino}

When the right-handed neutrinos have a mass above the electroweak scale, they can quickly decay into an electroweak gauge boson and a lepton doublet.  We will return to the case of lighter right-handed neutrinos in Sec.~\ref{sec:DL}.

The right-handed sneutrino also has a simple cosmology.  It can decay to a Higgsino and a lepton via the Yukawa coupling of Eq.~(\ref{eq:Yukawa}).  This Yukawa coupling is related to the neutrino masses as $y_N=\sqrt{m_{\nu} m_N}/v$. The decay will occur before the freeze-out of the LSP, so the sneutrino does not appreciably modify the cosmology of other supersymmetric relics in this case.

\subsubsection{Light axino}
 \label{sec:lightaxino}
 In Sec.~\ref{sec:CRaxino} we found that axino masses $m_{\tilde{a}} >$ TeV can be consistent with cosmology, but for $m_{\tilde{a}}$ less than the other superpartner masses, the axino was strongly constrained unless its mass was below 10 eV.  Here, we discuss this possibility in more detail.  We find that it requires a non-trivial spectrum for the superparntners, and also perhaps some non-minimal theoretical ingredients.

A very light axino has the potential to avoid constraints from a too-large dark matter abundance that may moreover be too warm. But as we will see below, such a light axino will only be possible for some values of $m_{S}$.  To understand this, we first review the origin of the axino mass, which requires a review of the structure of the one-field model.  It is in this model that such a light axino may be possible, 
since it avoids a Dirac mass for the axino that can be produced via a supergravity effect.  

In the one-field model, the saxion vacuum expectation value is induced by dimensional transmutation: quantum corrections to the saxion mass from its interaction with other fields can drive its mass-squared negative.  We call the fields to which the PQ field couples that are responsible for the dimensional transmutation $\psi$ and $\bar{\psi}$, with
\begin{equation}
W \ni y_\psi P \psi \bar{\psi}.
\end{equation}
In the KSVZ case, the most minimal option would be to identify these $\psi$ with the KSVZ quarks themselves.
The condition for successful dimensional transmutation on $m_S$ is set by 
\begin{equation}
m_S^2 \simeq \frac{y_\psi^2 N_\psi}{4\pi^2} m_{\tilde{\psi}}^2,
\label{eq:DT}
\end{equation}
where the saxion mass is defined around the vacuum rather than a large field value of $S$.

But this coupling can induce an axino mass.
A contribution to the axino mass comes from a loop with $\psi$'s and  their superpartners $\tilde{\psi}$'s. Assuming $N_\psi$ pairs of $\psi$ and $\tilde\psi$, the result of this loop is
\be
m_{\tilde{a}} \simeq \frac{y_\psi^2 A_{y_\psi}}{8\pi^2} N_\psi.
\ee

It remains to discuss the size of $A_{y_\psi}$ and $\tilde{m}_{\psi}$. If the $\psi$ fields are charged under gauge symmetries (as would be expected in the minimal KSVZ case), then their superpartners will receive a gauge-mediated contribution to their mass, and the $A_{y_\psi}$ is expected to be suppressed by a loop factor relative to $m_{\tilde{\psi}}$, $A_{y_\psi} \sim \frac{m_{\tilde{\psi}}}{16\pi^2}$, where we have set SM gauge couplings to be order unity. 
Therefore,
\be
m_{\tilde{a}} \sim \frac{y_\psi^2 m_{\tilde{\psi}} N_\psi}{(16\pi^2)^2} .
\label{eq:axino_mass}
\ee
Using Eq.~(\ref{eq:DT}), the relation between the masses of the saxion and axino is
\be
m_{\tilde{a}} \sim \frac{m_S^2}{16\pi^2 m_{\tilde{\psi}}} .
\ee
Rearranging, we find
\be
\label{eq:mSaxionLightAxino}
m_S \sim 4 \pi \sqrt{m_{\tilde{a}} m_{\tilde{\psi}}}.
\ee
Assuming $m_{\tilde{\psi}}$ is around the MSSM soft masses, i.e., $m_{\tilde{\psi}} \sim$ 10 TeV, and imposing the requirement $m_{\tilde{a}}< 10$ eV, we get $m_S \lesssim 0.1\ \rm GeV$.   For this reason, we have denoted the region below $m_{S} = 0.1$ GeV in Figs.~\ref{fig:KSVZ} and~\ref{fig:DFSZ} as the ``light axino" region.

In addition to the corrections from $\psi$, there can be quantum corrections from the field $Q$ responsible for the thermalization of the rotation.
Since $Q$ has a gauge charge and a Yukawa coupling $y_Q$ to the field $P$, we expect quantum corrections in Eq.~(\ref{eq:axino_mass}) 
with $\psi$ replaced with $Q$, which
requires $y_Q < 5 \times 10^{-4} ({\rm TeV}/m_{\tilde{Q}})^{1/2}$. 
We find that this bound on the Yukawa coupling is weaker than the requirement used in thermalization constraints, $y_Q S_{\rm th} \le T_{\rm th}$. 
Also, in the DFSZ model, there is another contribution to the axino mass generated at tree level via  axino-Higgsino mixing. This results in a contribution of size $\sim \mu v^2 / (\tan \beta f_a^2)$. This is well below $10$ eV, so this tree level contribution does not cause the axino to be heavier than desired.

In the DFSZ model or a non-minimal KSVZ model (where $\psi$ are the KSVZ quarks but more directly couple to the SUSY breaking, or new gauge-singlets $\psi$ distinct from the KSVZ quarks are introduced), the soft mass of $\psi$ can be larger than the MSSM soft masses of $10$~TeV and/or the $A_{y_\psi}$ can be more suppressed.  These model variations can then decrease the size of the axino mass induced from a given $m_S$. In an extreme case,  the $\psi$ fields responsible for dimensional transmutation could be gauge singlets and have 
additional couplings to the SUSY-breaking sector
that are $R$-symmetric. Then the dominant contribution to the $A$-term is from anomaly mediation.  
In this case, the axino mass is given by 
\begin{equation}
    m_{\tilde{a}}= \frac{y_\psi^4}{(16 \pi^{2})^2} m_{3/2}\sim \frac{m_{S}^4}{ m_{\tilde{\psi}}^4 (16 \pi^{2})^2} m_{3/2},
\end{equation}
where $m_{\tilde{\psi}}$ can be as high as $\sqrt{m_S f_a}$ without spoiling dimension transmutation, which requires $m_{\tilde{\psi}} < y_\psi f_a$. In summary, there are several possibilities to avoid the large axino mass, even if $m_S > 0.1$~GeV. 

On the other hand, it is an interesting model-building question to understand how a sub-GeV $m_{S}$ can be made consistent with the requirement of the visible superpartner masses being above a TeV. This could happen in a gauge-mediated model, with $m_{S} \sim m_{3/2}$.
Since the right-handed neutrinos are SM gauge neutral, the soft masses of right-handed sneutrinos would be much below the weak scale. This results in a nearly degenerate right-handed (RH) neutrino and sneutrino. This need not be problematic: the sneutrino can decay rapidly to a lepton and Higgsino as long as $m_{N}$ is above the TeV scale. 

The constraints that arise from LOSP decays to gravitinos may be absent in the case where the LOSP can decay into a light axino before BBN. For example, in the KSVZ model with a bino-like LOSP, the decay of the bino into the axino occurs before BBN begins if $f_a < 10^{12} (M_1/{\rm TeV})^{3/2}$ GeV, which is satisfied in all of the parameter space relevant for the kinetic misalignment mechanism. This scenario requires the axino mass below the LOSP mass, and as discussed above, this scenario requires $m_{\tilde{a}} < 10$ eV. 

Suppose that a model of supersymmetry breaking is constructed that allows the axino to be light. In this case, we would expect the gravitino to be heavier than the axino. We must therefore discuss constraints arising from an unstable gravitino. The gravitino would decay via Planck suppressed operators to the axino and axion with rate \cite{Hamaguchi:2017ihw}
\be
\Gamma_{\tilde{G} \rightarrow a \tilde{a}} = \frac{m_{3/2}^3}{192 \pi M_{Pl}^2}.
\ee

\begin{figure}[t!]
\includegraphics[width=4in]{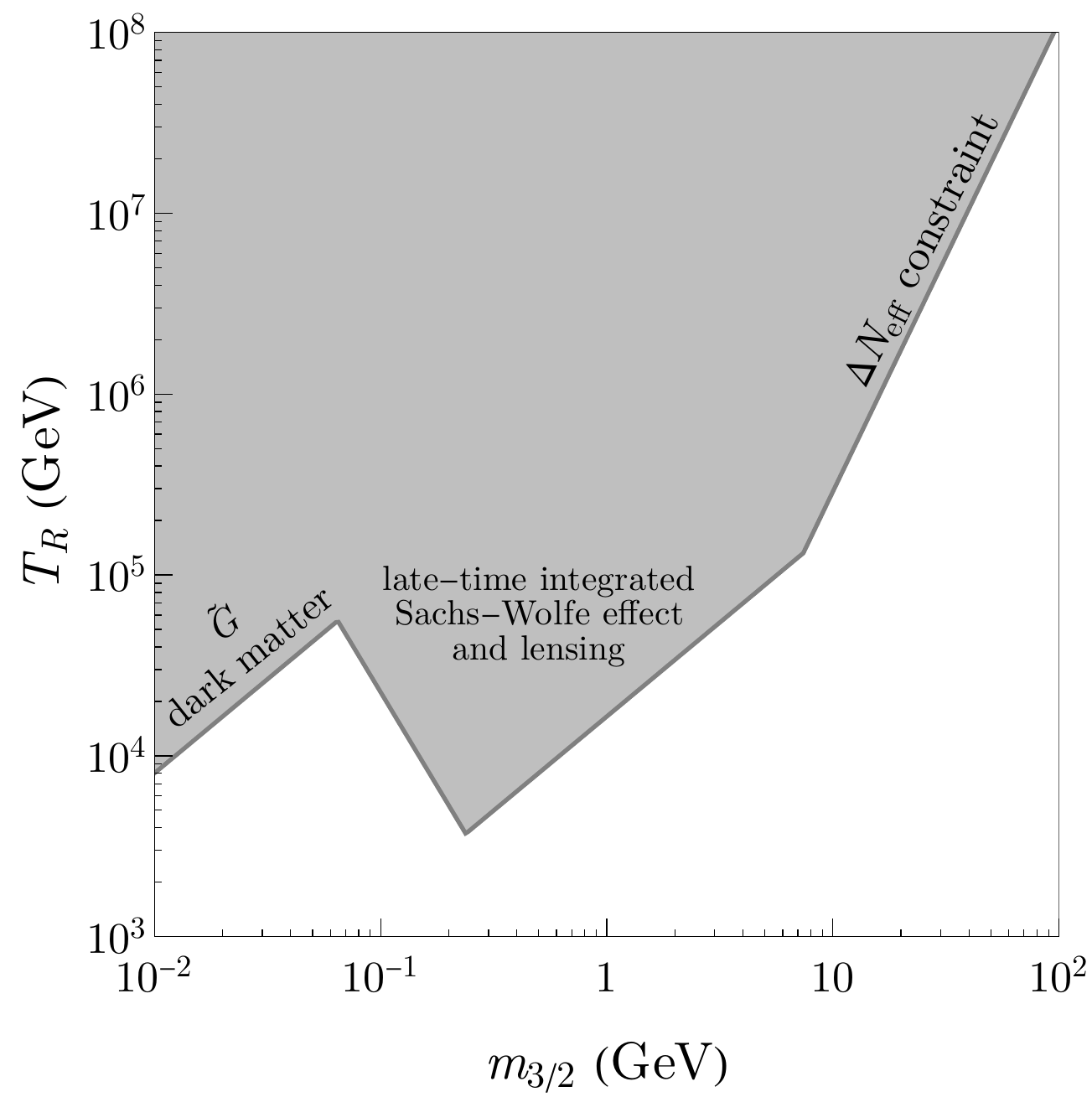}
    \caption{The bound on $T_R$ in the case where the gravitino can decay to an axino and an axion, $\tilde{G} \rightarrow \tilde{a} a$, relevant in the case where the axino is light, see Sec.~\ref{sec:lightaxino}. Here the gluino mass $M_3$ at the TeV scale is 5 TeV; the upper bound on $T_R$ approximately scales as  $M_3^{-2}$. }
    \label{fig:TRHgravitinoToaxino}
\end{figure}

The constraints arising from the decays of the unstable gravitino depends on its mass, and are summarized in Fig.~\ref{fig:TRHgravitinoToaxino}.  The figure is produced by using the gravitino production rate from Ref.~\cite{Kawasaki:2008qe} and the upper bound on the abundance of decaying dark matter from Ref.~\cite{Nygaard:2020sow}, see Ref.~\cite{Poulin:2016nat} for earlier work on decaying dark matter before {\it Planck} 2018.  For gravitino masses above $10$ GeV, the decays are before recombination, and the constraints come from the modification of the CMB through extra relativistic degree of freedom (at the largest masses, this can be summarized as an modification of $\Delta N_{\rm eff}$). If the gravitino is lighter than $10$ GeV, the dominant constraint comes from the effect of modifying the late-time integrated Sachs-Wolfe effect and weak gravitational lensing. 
For $m_{3/2} >0.2$ GeV, the decay occurs before dark-energy domination, and since the gravitational potential remains constant during matter domination, the constraint on the fraction of decaying dark matter does not depend on when exactly the decay occurs.
As $m_{3/2}$ gets even smaller, the bound on the fraction depends on when the decay occurs, and the abundance of the gravitino can be larger. For $m_{3/2} \lesssim 0.06$ GeV,
the gravitino abundance may be as large as the observed dark matter abundance, which places a constraint on $T_R$. In the case of a light axino (which typically requires $m_S<0.1$ GeV, see Eq.~(\ref{eq:mSaxionLightAxino})),  for $m_S\sim m_{3/2}$,
$T_R \lesssim 10^{5} \GeV$.  Note,  the right-handed neutrino masses in the freeze-out scenario may be of order 10 TeV.  So, it is possible that gravitino masses $m_{3/2}$ of order 0.1 GeV may allow reheat temperatures in excess of the right-handed neutrino mass, but if not,  Fig.~\ref{fig:DFSZ_TR} would apply rather than Fig.~\ref{fig:DFSZ}.

If one wanted to  avoid the strongest bounds on $T_R$ (by taking $m_{3/2}> {\mathcal O}(10)$ GeV), then $m_{3/2} \gg m_{S}$, and so a mild sequestering of $P$ from the supersymmetry-breaking sector would be required to keep $m_{S} \sim 0.1$ GeV.%
\footnote{
If the $A$ term in Eq.~\eqref{eq:VAterm} is much larger than $m_S$, $P$ will be trapped at a minimum at a large field value created by the $A$ term~\cite{Kawasaki:2006yb}. The supergravity effect is expected to generate $A\sim m_{3/2}$, so $m_{3/2} \gg m_S$ requires the suppression of the $A$ term or a potential that differs from Eq.~\eqref{eq:VAterm}.  
}

In this case, the gravitino could also decay to saxion and axino.  This decay can potentially be more problematic, depending on the decay profile of the saxion.  Indeed, visible decays of the saxion could disturb BBN or the CMB.  In a KSVZ model, the decay of a light saxion will primarily be to axions, with a small branching ratio to photons.  The small branching ratio mitigates the destructive power of these decays, and so the presence of saxions in gravitino decays are likely not problematic.
For DFSZ models, on the other hand, the saxion will undergo mixing with the Higgs boson. For saxions in the 0.1 GeV range, this will induce  saxion decays to electrons with a rate
\begin{equation}
\Gamma_{S\rightarrow e^+e^-} \simeq \frac{1}{8\pi} \frac{\mu^4 m_e^2 m_S}{f_a^2 m_h^4} \simeq \frac{m_e^2 \mu^4}{m_S^2 m_h^4} \times \Gamma_{S \rightarrow 2a},
\end{equation}
which is typically larger than $\Gamma_{S\rightarrow 2a}$. The BBN bound is then stronger than $\Delta N_{\rm eff}$ bound and requires $m_{3/2}Y_{3/2} < 10^{-14}$--$10^{-12}$ GeV for $m_{3/2}=100-1000$ GeV~\cite{Kawasaki:2017bqm}. The corresponding upper bound on $T_R$ is
\begin{equation}
T_R <  5\times 10^3 \GeV 
\left( \frac{m_{3/2}}{100~{\rm GeV}} \right) 
\left(\frac{3~{\rm TeV}}{m_{\tilde{g}}}\right)^2
\left( \frac{m_{3/2}Y_{3/2}}{10^{-14} \GeV} \right),
\end{equation}
where we assume that the gravitino is lighter than MSSM particles. The upper bound on $T_R$ is stronger the bound on it for $m_{3/2}\lesssim m_S$, so having $m_S\ll m_{3/2}$ does not help relaxing the bound on $T_R$ for the DFSZ model.

\subsubsection{Relic summary}\label{sec:relicssummary}

To summarize, the following scenarios are consistent with the BBN,  dark radiation, and (hot) dark matter constraints,
\begin{itemize}
    \item All of supersymmetric particles and the saxion have similar masses, and the LSP is a MSSM particle or the gravitino.
    \item The axino mass is below $10$ eV. This is possible only for the one-field model. In this case, $m_{S}< 0.1$ GeV in the minimal KSVZ model, but this constraint may be relaxed if there are new fields responsible for dimensional transmutation.
\end{itemize}

\section{Dirac lepto-axiogenesis: $m_N$ below the weak scale}
\label{sec:DL}

When the right-handed neutrino mass is below the electroweak scale, the Yukawa coupling for neutrinos will be small. The generation of the asymmetry can be in the ``freeze-in" regime and will be suppressed by the factor $\Gamma_N/H$, with $\Gamma_N$ the production rate for the $N$ asymmetry. As a result of this freeze-in process, a $B-L$ asymmetry is generated for $N$, and an (opposite) asymmetry is generated for the SM particles. Above the electroweak phase transition, sphalerons act on the SM portion of the asymmetry, converting an ${\mathcal O}(1)$ fraction of the $L$ asymmetry to baryon number.
And while the opposite asymmetry is  temporarily stored in the $N$, when they eventually decay, they do so symmetrically. At the epoch of decay the Majorana mass is important, and any sense of asymmetry in the $N$ is lost. So this would-be compensating asymmetry is washed out, and only the asymmetry from the portion stored in the SM particles remains. 
The asymmetry stored in $N$ could be also transferred to SM particles via lepton-number conserving scattering. As long as the scattering is ineffective before the electroweak phase transition, only the lepton asymmetry is partially washed out and the baryon asymmetry remains intact.

Although in this work, we assume the right-handed neutrinos have Majorana masses, there can in principle be a similar contribution when the SM neutrinos have Dirac masses. In this case, however, the scattering via the tiny Yukawa coupling of the neutrino ($y_{\nu} \sim 10^{-12}$) is too small to explain the observed baryon asymmetry. Ref.~\cite{Chakraborty:2021fkp} overcomes this difficulty by making the right-handed neutrino composite, and scattering with  right-handed neutrino sector can be effective above the compositeness scale. Ref.~\cite{Berbig:2023uzs} considers a direct coupling of an axion-like particle with the neutrino mass sector as well as more fermions so that the scattering can be more effective.
 
 In this section, we analyze this ``freeze-in" regime, which we refer to as Dirac Lepto-Axiogenesis, in analogy to Refs.~\cite{Dick:1999je,Murayama:2002je}, where the lepton asymmetry produced by the decay of a heavy field is initially stored with an equal and opposite contribution in the $N$ and $L$ before an epoch of left-right equilibration.

\subsection{Production of the asymmetry}
\label{sec:DLProd}

In the high $T_R$ limit, the asymmetry produced per Hubble time is given by
\begin{align}
\label{eq:YBminuL_highTR_lowmN}
    Y_{B-L} = \frac{n_{B-L}}{s} = \frac{c_{B-L} \dot\theta T^2}{s}  \frac{\Gamma_N}{H}.
\end{align}
When the expansion of the Universe is radiation dominated ($H \propto T^2)$, noting $\Gamma_{N} \propto T$ from Eq.~(\ref{eq:scatteringRate}),  we have $Y_{B-L} \propto T^{-2}$ and $Y_{B-L} \propto T$ for $T > T_S$ and $T < T_S$, respectively. This implies the production peaks at $T_S$. 
The value of $T_S$ can be determined by requiring that the axion reproduce the dark matter from the kinetic misalignment mechanism, see Eq.~(\ref{eq:TS_KMM}).  The rate is proportional to $y_N^2$, which can be traded for $m_{\nu} m_N/v^2$.

We find the produced baryon asymmetry to be
\begin{align}
\label{eq:YB_highTR_lowmN}
    Y_{B} \simeq 9 \times 10^{-11}
    \left( \frac{c_{B-L}}{0.1} \right) 
    \left( \frac{N_{\rm DW} m_S}{100 \GeV} \right)^{\scalebox{1.01}{$\frac{1}{3}$} }
    \left( \frac{10^9 \GeV}{f_a} \right)^{ \scalebox{1.01}{$\frac{2}{3}$} }
    \left( \frac{m_N}{10 \GeV} \right)
    \left( \frac{m_\nu}{0.05 \eV} \right)
    \left( \frac{g_{\rm MSSM}}{g_*(T_S)} \right)^{ \scalebox{1.01}{$\frac{5}{6}$} } .
\end{align}
It turns out that there is a relatively small range of $m_{N}$ of interest, approximately between 1 and 10 GeV.   For $m_{N}$ too small, $N$ decays can disturb BBN; see Sec.~\ref{sec:DL_CR_RHN}. And for $m_N$ too large, there is a danger that sphalerons can washout the asymmetry; see Sec.~\ref{sec:DLWashout}. The final baryon asymmetry will be suppressed by a washout factor $e^{-\Gamma_{\rm wo}t}$. It remains to do a robust calculation of $c_{B-L}$. In Appendix \ref{sec:AppendixFI}, we present a more detailed calculation of the freeze-in process, accounting for both scattering processes as well as decay processes that can all contribute to the generation of $N$.  The most important scattering processes that generate a $B-L$ asymmetry are $\ell_i+W \rightarrow N_j+H$, $\bar{H}+\ell_i \rightarrow N_j+W$, $\bar{H}+W \rightarrow N_j+\bar{\ell}_i$, $\ell_i+t \rightarrow N_j+q_3$, $\ell_i+\bar{q}_3 \rightarrow N_j+\bar{t}$, and $t+\bar{q}_3 \rightarrow N_j+\bar{\ell}_i$.  Here, $\ell_i$ is the lepton doublet in generation $i$, $N_j$ is the right-handed neutrino in generation $j$, $W$ is an $SU(2)_L$ gauge boson, $t$ is the right-handed component of the top quark, $q_3$ is the quark doublet in the third generation, and $H$ is the Higgs doublet. The time reversal of each process does not contribute significantly because right-handed neutrinos are not in thermal equilibrium with the bath, so their number density is negligible. Because the production peaks at $T_{S}$, which typically exceeds the supersymmetry scale, we also include processes involving superpartners.  The dominant contributions from decays are of the form $H_{u} \rightarrow \ell^{\dagger} N^{\dagger}$, and their supersymmetric analogs.  We find that these decays are generally subdominant to the scattering processes.

In computing the final baryon asymmetry, we integrate over time the $B-L$ production rate $\dot{n}_{B-L}$, which is obtained in Appendix \ref{sec:AppendixFI}. The black curves in  Fig.~\ref{fig:KSVZ}  (KSVZ) and in Fig.~\ref{fig:DFSZ} (DFSZ) with $n=1$ ($n=2$) in the left (right) panel, show the minimum $m_S$ necessary to reproduce the observed baryon asymmetry. These curves are obtained using $m_N = 11 \GeV$, which gives the most efficient baryon asymmetry production accounting for  washout as discussed later in Sec.~\ref{sec:DLWashout}. Larger $m_N$ leads to strong washout suppression, while smaller $m_N$ suppresses the production rate owing to the smaller Yukawa coupling. To the right of the kination line, the energy of the rotation dominates as determined from the KMM. This enhances $H(T_S)$ compared to the radiation-dominated case and further suppresses the freeze-in factor $\Gamma_N / H$ in Eq.~(\ref{eq:YBminuL_highTR_lowmN}) so that larger $\dot{\theta}$ and hence $m_S$ are required.  This explains why the black curve bends to the right. The Peccei-Quinn field fails to thermalize prior to $T_{\rm RM}$ in Eq.~(\ref{eq:TRM}) in the green shaded regions (or the regions above the green dashed curves for solely Higgs interactions), so the saxion thermalization creates entropy. In this case, as discussed below Eq.~(\ref{eq:thermTRM}), the viable parameter space collapses onto the negatively-sloped green boundary. The baryon asymmetry derived in this section is unaffected because the entropy is only produced prior to the dominant production of the asymmetry, i.e., $T_{\rm th} > T_S$.
As can be seen from the figures, the combination of freeze-in production and saxion domination is only possible with high-scale $m_S$.

If $T_R < T_S$ instead,
we find that the production per Hubble time given in Eq.~(\ref{eq:YBminusL_lowTR}) peaks at $T_R$. This differs from the discussion around Eq.~(\ref{eq:YBminusL_lowTR}) because $m_N < T_R$ in the freeze-in regime. The baryon asymmetry is diluted by entropy production and is given by
\begin{align}
\label{eq:YB_lowTR_lowmN}
    Y_{B} \simeq 10^{-10}
    \left( \frac{c_{B-L}}{0.1} \right) 
    \left( \frac{T_R}{10^5 \GeV} \right)
    \left( \frac{10^8 \GeV}{f_a} \right)
    \left( \frac{m_N}{3 \GeV} \right)
    \left( \frac{m_\nu}{0.05 \eV} \right)
    \left( \frac{g_{\rm MSSM}}{g_*(T_R)} \right)^{ \scalebox{1.01}{$\frac{1}{2}$} } .
\end{align}
For an unstable gravitino with $m_{3/2}\sim$ TeV $(\sim m_S)$, the BBN constraint requires $T_R\lesssim 10^5$ GeV $< T_S$, and this formula would apply.

\subsection{Washout}
\label{sec:DLWashout}
If scattering processes converting $\ell$ to $N$ are active while sphalerons are still in equilibrium, this can lead to efficient washout of the baryon asymmetry, i.e., too small $Y_B.$  Because in this freeze-in case, $N$ is light with mass below the weak scale, processes with $N$ in the final state are a possibility.

It is a quantitative question to determine whether washout is efficient.  We will see below that whether or not there is strong washout depends on the mass of the right-handed neutrino $m_{N}$.  We will focus on the dominant  $\ell + SM \rightarrow SM^{\prime}+ N$ scattering process, mediated by $W$-exchange, and determine what is the maximum value for $m_N$ for which this process is in equilibrium when sphalerons decouple. Because this process has a $t$-channel power IR singularity, care must be taken with the re-summation of the $W$-propagator.  This was done in Ref.~\cite{Ghiglieri:2016xye}, which also considered the question of washout in theories with light right-handed neutrinos (and the related question of right-handed neutrino production and equilibration).   We will ultimately adapt their numerical results to the present case, but we first sketch a simpler calculation that can give insight into the underlying physics and also give a reasonable approximation of the result.

Around the temperature at which sphaleron processes decouple, $T_{\rm sph}=130$ GeV \cite{DOnofrio:2014rug}, the dominant contribution to the scattering process can be thought of as ``indirect"; that is, it is a process where the interaction of the right-handed neutrino arises via left-right mixing induced by the Higgs boson vacuum expectation value.  Because of this, the washout rate can be found by considering the interaction rate for active neutrinos, and then accounting for this mixing.  

At finite temperature, the mixing angle is given by \cite{Ghiglieri:2016xye}
\begin{equation}
\label{eq:finiteTangle}
\theta^2 \simeq \frac{y_N^2 v(T)^2 M^2}{(M^2-m_{\ell}^2)^2 + (k_{0} \Gamma_{\nu})^2}.
\end{equation}
Here, the (asymptotic) finite-temperature mass squared for the lepton doublet is $m_{\ell}^2 =3g^2/16 T^2$; this mass differs by a factor of $\sqrt{2}$ when compared with the mass considered for phase transitions, see \cite{Anisimov:2010gy}.  We have emphasized the temperature-dependent nature of $v$ with a normalization $v(T=0)\simeq 173$ GeV, and included a term $k_0 \Gamma_{\nu}$ in the denominator corresponding to the thermal width of the active neutrino.  We  find $\Gamma_{\nu}$ by computing the cross section for scattering off of (anti)-quarks $\nu \barparen{q} \rightarrow \ell \barparen{q^\prime}$, which receives an enhancement both due to the $t$-channel exchange of a $W$-boson, and also the color multiplicity of the quark.  It is straightforward to compute and thermally average this cross section using Maxwell–Boltzmann statistics for the particles in the bath.  The RH neutrino scattering rate can then be approximated by  $\Gamma_{\rm sc} \simeq \theta^2 \Gamma_{\nu}$.  Note that the dominant process contributing to washout is  $\ell + SM \rightarrow SM^{\prime}+ N$ (owing to the full thermal abdundnace of the initial state particles); it is related by crossing symmetry (and the small mixing angle) to $\Gamma_{\nu}$.  

When calculating $\Gamma_{\nu}$ we retain the $W$ boson mass (to regulate the $t$-channel divergence that might otherwise be  present). For its mass, we  take into account not only the temperature dependence of the Higgs expectation value, i.e., 
 $m_{W}^2= g^2 v(T)^2/2$, but also a Debye mass.  So, we have
$m_{\tilde{W}}^2 = m_{W}^2 + m_{W,{\rm Debye}}^2$, with
\begin{equation}
    m_{W,{\rm Debye}}^2=\left(\frac{2}{3} + \frac{n_{S}}{6}+\frac{n_{G}}{3}\right) g^2 T^2.
\end{equation}
Here, the number of scalar Higgs doublets $n_{S}=1$ and the number of generations $n_{G}=3$. 

Setting the value for $\Gamma_{\rm sc}$ equal to the Hubble expansion rate at $T=130$ GeV allows us to find an upper bound on the mixing angle such that this process is not in equilibrium.  Trading $y_N^2 = m_{N} m_{\nu}/ v^2$  allows us to translate this into a bound on the maximum value of $m_{N}$ that avoids washout. Here, we use $m_{\nu} = 120$ meV, a value that effectively corresponds to the maximum sum of the neutrino masses. Using this simplified approach, we find that washout is avoided for $m_{N} \lesssim 10$ GeV.

This is consistent with more precise results computed in~\cite{Ghiglieri:2016xye}.  
To get a more precise determination of the maximum allowed $m_N$, we interpolate their results in the region of interest.
For masses near 10 GeV, their displayed washout rate can be approximated by
\begin{equation}
\left. \frac{\Gamma_{\rm wo}}{H} \right|_{T=130\GeV} \simeq 0.6 
\left(\frac{m_N}{8 \GeV}\right)^{ \scalebox{1.01}{$\frac{3}{2}$} } 
\left( \frac{m_\nu}{8.7~{\rm meV}} \right) \, .
\label{eq:WOoverH}
\end{equation}
It turns out that the baryon asymmetry is generated more efficiently for the largest possible values of the active neutrino masses.  The reason is that the value of $B-L$ is  proportional to $y_N^2 \propto m_N m_\nu$ while the maximal $m_N$ avoiding washout is proportional to $m_\nu^{-2/3}$.  Saturating the bound $\sum m_\nu \leq 0.12 \eV$~\cite{Planck:2018vyg} and accordingly rescaling the washout rate, we find that $m_N \simeq 11 \GeV$ leads to the most efficient production of $Y_{B-L}$.  This is determined by  maximizing $Y_{B-L} \propto m_N e^{-\Gamma_{\rm wo}t}$ and using Eq.~(\ref{eq:WOoverH}); we take $t = 1/(2H)$ since the universe is radiation dominated at the electroweak phase transition (kination is predicted by the KMM to end by $T \simeq 2 \times 10^6 \GeV$~\cite{Co:2021lkc} for the QCD axion). We use the value $m_N = 11 \GeV$ in Figs.~\ref{fig:KSVZ} and~\ref{fig:DFSZ} to show the lowest $m_S$ consistent with the observed baryon asymmetry, where Eq.~(\ref{eq:YB_highTR_lowmN}) along with the washout suppression factor $e^{-\Gamma_{\rm wo}t}$ is used to compute $Y_B$. For smaller $m_N$ with fixed $m_\nu$, the $B-L$ production becomes less efficient due to a smaller Yukawa coupling and the prediction on $m_S$ becomes larger. For larger $m_N$, the washout of $B-L$ becomes very efficient and the prediction on $m_S$ becomes exponentially larger.

\subsection{Cosmological relics}
\label{sec:DLCR}

In the freeze-in scenario, $m_N$ is required to be below the electroweak scale. We discuss extra constraints arising from such a light right-handed neutrino as well as whether or not new possible spectra can result. 

\subsubsection{Right-handed neutrino}
\label{sec:DL_CR_RHN}
  The washout discussion above sets an upper limit on the right-handed neutrino mass, which means that they will decay less rapidly than in the freeze-out case.  If they are too light, then the right-handed neutrinos can disturb BBN upon decay.  This  limits $m_{N}>2$ GeV \cite{Abdullahi:2022jlv}.  Accounting for both the bound from BBN and the washout discussion, we are interested in the range $  2 {\rm \; GeV} <  m_{N} < {\mathcal O} (10)$ GeV.  

\subsubsection{Right-handed sneutrino}\label{sec:DLCR_snu}
 When the right-handed sneutrino has a mass above the MSSM superpartners, its cosmology remains simple.
The decay rate of the right-handed sneutrino into the Higgsino and the left-handed neutrino is
\begin{align}
\Gamma_{\tilde N, {\rm dec}} \simeq \frac{1}{8\pi} \frac{m_\nu m_N m_{\tilde{N}}}{v^2}.
\end{align}
The decay occurs before the annihilation of the Higgsino occurs even for the smallest value of $m_N$  we consider ($m_{N}=2 \GeV$). The large $m_{\tilde{N}}$ can be readily the case for gravity mediation. In gauge mediated models of supersymmetry breaking, where the mass of the saxion can be below $0.1$ GeV to keep the axino light, the right-handed sneutrino mass may be below the MSSM superpartner masses. On the other hand, if we gauge $B-L$, take $g_{B-L}$ sufficiently large,  and the messengers have $B-L$ charge, then the right-handed sneutrino can get a mass from this gauge mediation, in which case it still may be heavier than the lightest MSSM particle. In this case, the right-handed stau will get a similar mass, so it is unlikely to be the LSP; the bino or Higgsino are likely to be the lightest MSSM particle. 

In the gauge mediation without $B-L$ and
in some parameter space in gravity mediation, 
the right-handed sneutrinos are lighter than the MSSM particles. This case, however, is not cosmologically viable. 
First, the right-handed sneutrino cannot be the lightest supersymmetric particle.  While it may not reach equilibrium, even its freeze-in abundance would be dangerous. The dominant contribution comes from production in association with other superpartners at temperatures of order the SUSY scale, where $\Gamma_{\tilde N}/H  \sim y_N^2 M_{\rm Pl}/m_{\rm SUSY}$.  This leads to a too-large abundance, so we can exclude this possibility.

The axino or gravitino can be lighter than the right-handed sneutrino, and the decay of the right-handed sneutrino into them might solve the overproduction problem mentioned above. However, we now discuss why this is not the case. Decays to the gravitino would proceed via a 1/$F$ operator, with $F$ the scale of supersymmetry breaking. But even for a TeV-scale sneutrino, ensuring that these decays occur prior to BBN requires a sub-GeV gravitino. Such a gravitino is not innocuous. Not only would the abundance of the gravitino resulting from the sneutrino decays be large, but thermal production of such a gravitino could itself be large.  Then, to avoid too much dark matter from this gravitino would require a light axino to which the gravitino could itself decay.  However, decays of the gravitino to the axino are Planck suppressed, and for this GeV-scale gravitino, these occur too late--after the CMB is established. Therefore, a sneutrino decaying to a gravitino does not seem to provide an acceptable cosmology.  The other possible decay,  $\tilde N \rightarrow \tilde{a} N^{(*)}$, is too slow to be consistent with BBN. Decays will be most rapid in the DFSZ case, where the sneutrino could decay as $\tilde{N} \rightarrow \tilde{a} \ell (h)$.   However, even in this most optimistic case, for the smallest $f_{a}$, the decay of the sneutrino to the axino is too slow.

In summary, a sneutrino L(O)SP seems problematic, even in the Dirac lepto-axiogenesis case, so the spectrum appears to require a right-handed sneutrino with mass above at least some of the MSSM superpartners.

\section{Domain walls and parametric resonance}
\label{sec:DWandPR}

The initial motion of $P$ is generically elliptical, not circular. Motion in both the radial and angular modes is generated by the curvature of the potential.
For sufficient ellipticity, parametric resonance (PR) may occur~\cite{Co:2020dya}. This can lead to the production of fluctuations of the $P$ field.  
Such fluctuations have amplitudes comparable to the zero-mode amplitude and so have the potential to restore the PQ symmetry~\cite{Tkachev:1995md,Kasuya:1996ns,Kasuya:1997ha,Kasuya:1998td,Tkachev:1998dc,Kasuya:1999hy}.  In this case, it may  generate domain walls at the QCD phase transition, which may be problematic.  A second concern is that axion fluctuations produced during PR may give an excessive contribution to $N_{\rm eff}$, or alternatively, excessively warm dark matter.  When PR  occurs at temperature $T_{\rm PR} \simeq T_S (S_{\rm PR} / f_a)^{2/3}$ with $S_{\rm PR}$ the saxion field value at PR, the axions are produced with momenta $k_a$ of order $0.1 m_S$. These relativistic axions cool by redshift, $k_a \propto g_*^{1/3} T$, and the experimental constraint on warm dark matter can be expressed as a bound on the axion velocity at $T = 1\eV$~\cite{Irsic:2017ixq, Lopez-Honorez:2017csg},
$ k_a|_{T = 1\eV}/m_{a} \lesssim 2\times 10^{-4}$. Together, this gives an upper bound on the saxion mass as
\begin{equation}
\label{eq:WDM}
    m_S \lesssim 10 \GeV 
    \left(\frac{N_{\rm DW}}{6}\right)^{\scalebox{1.01}{$\frac{1}{2}$} }
    \left(\frac{S_{\rm PR}}{30 f_a}\right)
    \left(\frac{2 \times 10^{10} \GeV}{f_a}\right) ,
\end{equation}
where $S_{\rm PR}$ differs significantly between the one- and two-field models as we discuss below.

In this section, we will discuss the issue of PR in more detail---both potential constraints on the parameter space, as well as ways in which these constraints might be avoided. The efficiency of PR depends on the form of the saxion potential, so differs for the one- and two-field models.  We analyze each of these cases in turn.

\begin{figure}
\includegraphics[width=0.495\linewidth]
{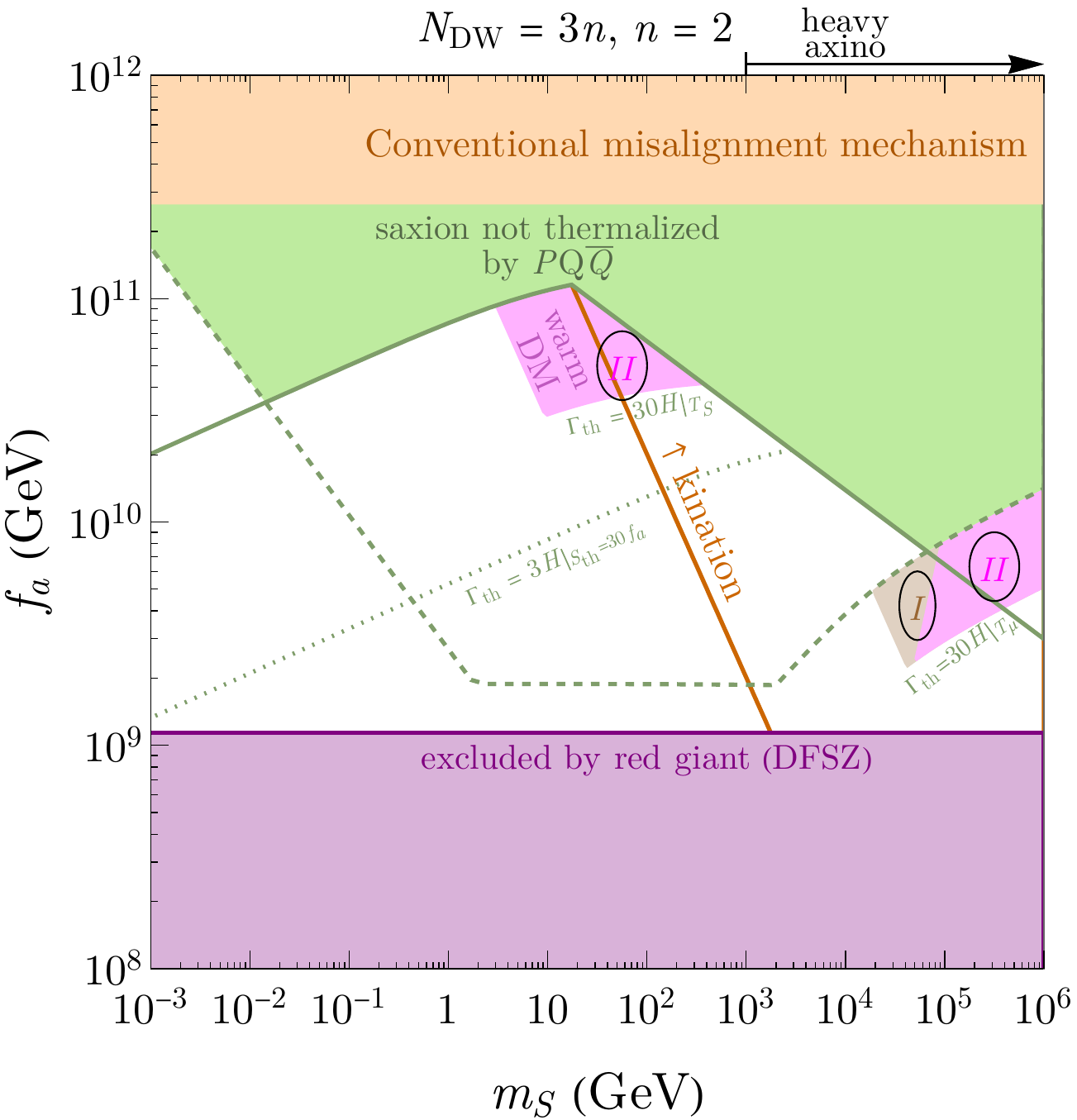} 
\caption{Additional warm dark matter constraints for a DFSZ axion with $n=2$ and $\mu = \max(m_S,2 \TeV)$.  Regions marked {\it I} correspond to the one-field model, whereas {\it II} corresponds to the two-field model. The dashed green line shows the boundary of the region below which thermalization of the saxion occurs via interactions with the Higgs multiplets.  The region below the solid green line can thermalize in the presence of new fields $Q$; however, thermalization occurs after PR and there can be a constraint from warm dark matter in the magenta region on the top.  The regions at the bottom right of the dashed green line thermalize via interactions with Higgs fields at $T_\mu$, but $T_\mu$ may not be before onset of parametric resonance, and constraints from warm dark matter may apply.  Since PR occurs earlier in the one-field model, the region is larger, and extends to the brown shaded region marked {\it I}. The magenta region marked {\it II} produces warm dark matter in both the one-field and the two-field case.}
\label{fig:WDM}
\end{figure}

\subsection{Two-Field Model}
\label{sec:PRTwoField}

In a two-field model, Eq.~(\ref{eq:two-field}), we do not anticipate a domain wall problem even in the presence of PR. In this case the PQ symmetry is not restored after PR, so any domain walls that result from fluctuations are sub-horizon in nature, which can safely annihilate for any value of $N_{\rm DW}$. PQ symmetry restoration is avoided because the complex fields are strongly fixed to the moduli space $P \bar P = v_{\rm PQ}^2$---the energy required to restore the PQ symmetry, $v_{\rm PQ}^4$, is far larger than that available in the PQ fields at PR.

Although domain walls are not a concern, PR could produce an excessive amount of dark radiation or warm axion dark matter. However, in this two-field model, it turns out to be relatively straightforward to avoid the occurrence of PR altogether. The key observation is that if thermalization occurs prior to the onset of PR, parametric resonance is avoided.
The thermalization process removes energy from the rotation while preserving its angular momentum, thus transforming the elliptical motion into a circular one~\cite{Co:2019wyp}, and PR is prevented. This requires a thermalization rate higher than the PR rate. The former depends on the particle content in the bath that couples to $P$, while the latter depends on the potential of $P$ in the radial direction. For the two-field model, it has been shown numerically in Ref.~\cite{Co:2020jtv} that PR occurs when $S$ decreases to $\mathcal{O} (30) f_a$ for an ellipticity of order unity. It remains to check whether thermalization options exist that are sufficiently efficient.

As discussed in Sec.~\ref{sec:review},  if there exists a coupling between $P$ and PQ-charged particles $Q$ and $\bar{Q}$,
the largest thermalization rate is $\Gamma_{\rm th} \lesssim b T^3/S^2$. The requirement that thermalization occurs prior to PR (which would occur when the saxion field value reaches $S_{\rm PR}$), leads to the constraints
\begin{align}
    N_{\rm DW} m_S & \gtrsim 150 \GeV \ N_{\rm DW}^{-1}
    \left( \frac{0.1}{b} \right)^3 
    \left(\frac{S_{\rm PR}}{30 f_a}\right)^4 
    \left(\frac{f_a}{10^{10} \GeV}\right)^5
    \left( \frac{g_*(T_{\rm th})}{g_{\rm MSSM}} \right)^{ \scalebox{1.01}{$\frac{5}{2}$} } \ \ && \textrm{for RD} , \label{eq:Sth_PR_RD} \\
    f_a & < 3 \times 10^{10} \GeV \left(\frac{b}{0.1}\right)^{ \scalebox{1.01}{$\frac{1}{2}$} }
    \left(\frac{30 f_a}{S_{\rm PR}}\right)^{ \scalebox{1.01}{$\frac{1}{2}$} }
    \left( \frac{g_{\rm MSSM}}{g_*(T_{\rm th})} \right)^{ \scalebox{1.01}{$\frac{1}{2}$} } \ \ && \textrm{for MD}.
\label{eq:Sth_PR_MD}
\end{align}
The first (second) line applies to the case where thermalization occurs during radiation (matter) domination. These respectively correspond to the segments of the green dotted curve in Fig.~\ref{fig:WDM} to the left (right) of the kination line. Points below this line allow thermalization at $S_{\rm th} = 30 f_a$, prior to PR. 
The first constraint shows thermalization is easier at larger $m_S$. This is because a larger $m_S$ allows the saxion to reach a given field value $S$ at a higher temperature.  Note that the thermalization rate  $\Gamma_{\rm th} =bT^{3}/S^2$ is enhanced relative to the expansion rate $\sim T^2/\MPl$ by the higher temperature at $S=S_{\rm PR}$. This observation turns out to be relevant for the one-field model discussed in the next subsection.

For the KSVZ axion, should the $Q$ and $\bar{Q}$ fail to enter the bath, thermalization should instead proceed via the interactions with gauge bosons that arise upon integrating out the PQ fermions.  In this case, the rate is given by $\Gamma_g = 10^{-5} N_{\rm DW}^2 T^3/S^2$.  Note that the scaling is the same as in the case discussed above, just with a smaller numerical prefactor, $b \rightarrow N_{\rm DW}^2 10^{-5}$. The corresponding constraints can be straightforwardly obtained from Eqs.~(\ref{eq:Sth_PR_RD}) and~(\ref{eq:Sth_PR_MD}).  In this case, PR could not be avoided for any of the allowed parameter space shown.

In the DFSZ model it is also possible that no new thermalizing fields (i.e.~the $Q$ and $\bar{Q}$ above) exist.  In this case, however, there is the possibility that thermalization could proceed via interactions of  $P$ with the Higgs fields with the rate in Eq.~\eqref{eq:Gamma_S_higgsino}.
In the relevant parameter space, $T \gg m_S$ so scattering dominates. 
For $n=2$, thermalization is most efficient at highest temperature, $T_\mu$, as illustrated in the right panel of Fig.~\ref{fig:DFSZTherm}. Requiring successful thermalization at $T_\mu$ results in a constraint on $f_a$ shown in Eq.~(\ref{eq:n2Therm}) and the horizontal and positively-sloped segments of the green dashed line in Fig.~\ref{fig:WDM}. In this two-field model, $T_\mu$ can be higher than the temperature at which PR occurs $T_{\rm PR} \simeq T_S (S_{\rm PR} / f_a)^{2/3} \simeq 10~T_S$, so thermalization prevents PR and the production of warm dark matter altogether. 
In Fig.~\ref{fig:WDM}, the magenta regions have too warm dark matter.  The physics described above---where PR warm dark matter is altogether avoided---sets the left boundary of the magenta region in Fig.~\ref{fig:WDM} at high $m_S$ labeled by {\it II}.  

In the regions above the green dotted line or at high $m_S$ where $\mu < T_{\rm PR}$, PR is not prevented and axion fluctuations are created, which may become warm dark matter. The constraint is given by Eq.~(\ref{eq:WDM}) and determines the left boundaries of the magenta region at large $f_{a}$ labeled by {\it II} and of the brown region labeled by {\it I}. 

Also, the axion and saxion still have a sizable mixing when the zero-mode rotation is not circular or significant fluctuations are present, so that the  axions produced in PR have the potential to be thermalized together with the saxion. However the precise efficiency of this process is uncertain.  When the rotation is sufficiently circular, the angular fluctuations become derivatively coupled, which suppresses the thermalization rate. This implies that a fraction of axion fluctuations may fail to thermalize and subsequently contribute to warm dark matter. A precise determination of the process is left for future work. However, for reference, we truncate the warmness constraint by assuming that a thermalization rate ten times larger than necessary to thermalize the saxion, $\Gamma_{\rm th} = 10 \times 3 H$, is sufficient to thermalize axion fluctuations down to 10\%. This determines the lower boundaries of the two magenta regions. One can rescale the constraint once the precise condition is established. For example, if $\Gamma_{\rm th} / 3H = 100$ turns out to be necessary, the lower magenta boundaries will further shift downward by the same factor as the current magenta boundaries shift from the green lines, i.e., $f_a \propto 3H/\Gamma_{\rm th}$.

\subsection{One-Field Model}
\label{sec:PROneField}

In contrast to the two-field case, avoiding PR entirely by early thermalization seems difficult in the one-field case.  The one-field model potential deviates more strongly from a purely quadratic potential; this implies a larger $P$ self-interaction and more efficient PR than the one-field model. Indeed, the numerical analyses performed in Ref.~\cite{Co:2020jtv} show PR occurs in the one-field model as early as when the radial field value reaches $S \sim 10^6 f_a$ even for $\epsilon = 0.5$, with $\epsilon$ the parameter introduced in Sec.~\ref{sec:Therm} that describes how circular the motion is ($\epsilon=1$  is purely circular). 
 This constraint on $m_S$ in Eq.~(\ref{eq:Sth_PR_RD}) with $S_{\rm PR} \sim 10^6 f_a$ is severe, so PR inevitably occurs and domain walls may form.

One possibility to avoid problematic domain walls is simply to assume $N_{\rm DW} = 1$.  
In this case, following the non-thermal restoration of the PQ symmetry, a cosmic string will form and allow the domain walls to decay. However, especially for the 
DFSZ case, this assumption is non-minimal, and would require the introduction of additional matter.\footnote{In the DFSZ case with the minimal field content $N_{\rm DW}=3n$ where $n$ appears in the superpotential as $P^n H_{u} H_{d}$.}
Another potential possibility present for $N_{\rm DW}>1$ is to allow for the presence of explicit PQ breaking.  Explicit breaking allows domain walls to become unstable and decay \cite{Sikivie:1982qv}. 
 Rapid enough decays require $f_a \lesssim 10^8 \GeV$ based on Refs.~\cite{Hiramatsu:2012sc, Kawasaki:2014sqa} assuming a phase of unity between the explicit breaking minimum and the PQ-conserving one. However, this is in tension with the astrophysical bound on $f_a \gtrsim 10^9 \GeV$ for DFSZ~\cite{Capozzi:2020cbu,Straniero:2020iyi}, so for DFSZ, this approach cannot solve the domain wall problem.
 
It may also be possible that the PQ symmetry is never non-thermally restored despite the large fluctuations generated at PR. This is because the large angular momentum in the field space gives a contribution to the effective potential that strongly disfavors the origin. Namely, the centrifugal force repels the rotating field from reaching the origin so fluctuations still center around a large radius. A numerical simulation is warranted to determine whether such dynamics is in fact realized, and we leave this for future work.

Even if domain walls are absent, just like in the two-field case, there is a potential concern that axions produced in parametric resonance could lead to too much dark radiation or excessively warm dark matter. However, precisely the same reason that PR is difficult to avoid in the one-field case makes this problem potentially less severe.  In particular, the early onset of parametric resonance $S_{\rm PR} \gg f_a$ in the one-field case means that the axion momenta undergo a larger redshift so the axions may be sufficiently cold to be dark matter. This is reflected in the upper bound on $m_S$ in Eq.~(\ref{eq:WDM}), where $S_{\rm PR} \gg f_a$ significantly relaxes the constraint. The left boundary of the brown region marked by {\it I} in Fig.~\ref{fig:WDM} is determined by Eq.~(\ref{eq:WDM}) for $S_{\rm PR} = 10^4 f_a$. This brown region extends to the magenta region, and the lower boundary is set in the same way (possible thermalization of axion fluctuations) as the magenta lower boundary elaborated in the previous subsection.

\subsection{Parametric resonance summary}

\begin{itemize}
\item
In the one-field model, PR becomes efficient much before the radial direction reaches the minimum. As a result, it is difficult to avoid PR, and the PR may restore the PQ symmetry and produce domain walls. However, it is possible that the centrifugal force arising from the rotation may prevent symmetry restoration, but demonstrating this would require additional study beyond the scope of this work. Even if symmetry is restored, the choice domain $N_{\rm DW}=1$ avoids the domain walls problem, although this requires non-minimal matter content for the DFSZ model.

\item
In the two-field model, PR becomes efficient only when the radial direction is close to the minimum.  Thermalization may occur before PR becomes efficient, so that the motion becomes circular and the resonance band disappears.  However, even if PR occurs the symmetry will not be restored and there is no domain wall problem.

\item
In other model, should PR occur, axions produced at this epoch can be warm. The produced axions, however, may be absorbed by the thermal bath through the mixing with the saxion. If thermalization is sufficiently efficient, the warm DM may be avoided; this condition constrains the parameter space in the $m_{S}$ vs.~$f_{a}$ plane.

\end{itemize}

\section{Discussion and summary}
\label{sec:conclusion}

In this paper we explored the cosmological scenario where the QCD axion field rotates in  field space, and the PQ charge of the rotation is transferred into the baryon asymmetry of the Universe via the right-handed neutrino sector. We took the masses of the right-handed neutrinos to be  small enough that they may be produced from the thermal bath. We focused on supersymmetric theories and discussed the required range of the saxion mass. We found that the saxion mass may be as light as $\mathcal{O}(10)$ MeV. This is in contrast to the case where the right-handed neutrino mass is large and the production of the baryon asymmetry occurs only via the dimension-five operator $(\ell H)^2$, for which the saxion mass should be $\mathcal{O}(10-1000) \TeV$~\cite{Barnes:2022ren}. Because a large saxion mass is not required, our scenario is consistent with a wide range of supersymmetry-breaking and mediation scenarios, including the gravity mediation with the gravitino mass around $\mathcal{O}(1)$ TeV and the gauge mediation with a small saxion mass.

The baryon asymmetry production can proceed in one of two ways. In one case, the right-handed neutrino is heavier than the weak scale, and its Yukawa interaction and the $B-L$ violation by the right-handed neutrino mass enter thermal equilibrium before the electroweak phase transition.  These decouple as the right-handed neutrinos become non-relativistic.
The final $B-L$ asymmetry is determined by freeze-out at a temperature around the right-handed neutrino mass. The right-handed neutrino mass should be above 10 TeV in this freeze-out case. The saxion mass may be as low as $10/100$ MeV for the DFSZ/KSVZ model, below which the saxion produces negative/positive $\Delta N_{\rm eff}$ excluded by CMB observations.

In another case, the right-handed neutrino is lighter than the weak scale, and its Yukawa interaction never enters into thermal equilibrium. A non-zero $B-L$ asymmetry of the SM particles is ``frozen-in"  via the Yukawa interaction. Although the summed $B-L$ asymmetry of the SM particles and the right-handed neutrino vanishes before the right-handed neutrino becomes non-relativistic, the sphaleron process is sensitive only to the $B-L$ asymmetry of the SM particles, and a non-zero baryon asymmetry is produced. This is similar to Dirac leptogenesis, so we  christened this scenario Dirac lepto-axiogenesis, although the right-handed and SM neutrinos are eventually Majorana fermions. The right-handed neutrino should be heavier than $\sim$ 1 GeV; below this value, the late-time decay of the right-handed neutrino disturbs BBN. The freeze-in production of the $B-L$ asymmetry is the most effective when the axion rotation reaches the minimal of the saxion potential. The saxion mass $m_S$ is determined as a function of $f_a$ and $m_N$. The lowest possible $m_S$ is achieved when the right-handed neutrino mass is 11 GeV. If heavier, the Yukawa interaction becomes effective before the electroweak phase transition and the $B-L$ asymmetry of the SM particle is washed out. To compensate for this, the saxion mass would need to be much larger than that for $m_N < 11$ GeV.  The minimum saxion mass is $\mathcal{O}(10/100)$ GeV for $f_a$ saturating the astrophysical lower bound in the DFSZ/KSVZ model and larger for larger $f_a$.  Thus,  this low $m_{N}$ scenario can accommodate supersymmetry near the weak scale.

While we have considered the case where we initiate the rotation of the PQ symmetry-breaking field via a PQ-breaking operator, we could instead consider initiating the rotation of other scalar fields $X$, such as the MSSM flat directions, via the Affleck-Dine mechanism.  It is possible that $X$ then transfers its charge (i.e., angular momentum in field space) to the PQ symmetry-breaking field~\cite{Domcke:2022wpb}. Such a scenario may be compatible with dynamical PQ-breaking scenarios~\cite{Choi:1985cb,Feldstein:2012bu,Harigaya:2015soa}, where the saxion has a steep potential and cannot take on a large field value.  Also, in this case, constraints from parametric resonance are avoided since the radial direction of the PQ-breaking field is never excited and the rotation of the PQ-breaking field is circular. The two-field dynamics is as follows. Almost all of the charge of the system is initially stored in $X$, and the large initial field value suppresses the transfer rate. But as the field value of $X$ becomes smaller, the charge transfer becomes efficient and the system reaches chemical equilibrium. 
If $m_S < m_X$, when the charge transfer becomes efficient, almost all of the charge is transferred into $P$, and the dynamics afterward is identical to the case where the rotation was simply initiated in $P$ itself. As long as the freeze-out or freeze-in production dominantly occurs after the transfer, the analysis in this paper is immediately applicable.
On the other hand, of $m_S > m_X$, 
even if the charge transfer becomes efficient,
most of the charge is still stored in $X$ as long as $|X|> f_a$. The angular velocity of the axion $\dot{\theta}$ at chemical equilibrium is as large as $m_X$. Once $|X|$ becomes as small as $f_a$,
most of the charge of $X$ will transferred to the axion rotation with radius $f_{a}$ and $\dot{\theta}\sim m_{X}$, after which $\dot{\theta}$ decreases in proportion to $T^3$. The saxion mass in our analysis can then be understood as $m_X$. If $X$ can also transfer its charge to the MSSM particles directly, the coefficient $c_{B-L}$ can be modified by an $\mathcal{O}(1)$ factor.

Although our main focus is on supersymmetric theories and the prediction of the possible range of the saxion mass, most of our analysis  also applies to non-supersymmetric theories with a wine-bottle potential. For example, consider the case where the asymmetry is created via the freeze-out at $m_N$.
When the freeze-out occurs after the axion rotation reaches the minimum of the saxion potential, just as in supersymmetric theories, $\dot{\theta}$ at freeze-out is independent of $m_S$ after fixing axion dark matter abundance by kinetic misalignment, so the relation between $f_a$ and $m_N$ is unchanged (modulo a minor modification of $c_{B-L}$). In particular, the horizontal segments of the contours in Figs.~\ref{fig:KSVZ} and \ref{fig:DFSZ} remain horizontal in non-supersymmetric theories. Should  freeze-out occur before the axion rotation reaches the minimum of the saxion potential, $\dot{\theta}$ at higher temperatures is not the same as the saxion mass around the minimum, so a qualitative modification is required. For example, the vertical segments of Figs.~\ref{fig:KSVZ} and \ref{fig:DFSZ} will have a finite slope $f_a \propto m_S^{2}$ in non-supersymmetric theories.
The freeze-in production dominantly occurs when the axion rotation reaches the minimum of the saxion potential, so $m_S$ in our figures for the freeze-in case can be simply interpreted as the mass of the saxion around the minimum of the wine-bottle potential. In both cases, the constraints from supersymmetric relics are lifted.

\section*{Acknowledgements}
This work is in part supported by the Department of Energy under grant number DE-SC0011842 at the University of Minnesota (RC), under grant number DE-SC0007859 (AP), by Grant-in-Aid for Scientific Research from the Ministry of Education, Culture, Sports, Science, and Technology (MEXT), Japan (20H01895), and by World Premier International Research Center Initiative (WPI), MEXT, Japan (Kavli IPMU) (KH). 

\appendix

\section{Computation of freeze-in production rate} \label{sec:AppendixFI}
In this Appendix we discuss the production of the $B-L$ asymmetry that occurs in the Dirac lepto-axiogenesis scenario in Sec.~\ref{sec:DL}.  The asymmetry is generated by an out-of-equilibrium (freeze-in) process.  

The $B-L$ asymmetry may be generated by both $2 \rightarrow 2$ scattering processes as well as two-body decays.  Because the right-handed neutrino is not in thermal equilibrium with the bath, its number density is suppressed, and processes with a right-handed neutrino in the initial state do not contribute significantly. We find the dominant production processes are $2 \rightarrow 2$ scattering processes, but decay processes make a non-negligible numerical contribution. The dominant scattering processes have a logarithmic rather than power law divergence, so we can find the approximate production rate without doing a full thermal field theory calculation that would include careful re-summation.

In our computation, we have used the SARAH package \cite{Staub:2008uz}  to evolve the couplings from the weak scale to the temperature scale where the dominant contribution to the baryon asymmetry occurs (typically $T_{S}$).  We evaluate coupling constants at the scale $\mu=2 \pi T$, the energy of the first Matsubara mode, as advocated in Ref.~\cite{Kajantie:1995dw}.

For a  $2 \rightarrow 2$ scattering process $(1 + 2 \rightarrow 3 + 4)$, the contribution to the rate of $B-L$ production is given by
\begin{equation}
\dot{n}_{B-L} \supset \frac{2(\mu_1+\mu_2)}{T} \int d\Pi_1 d\Pi_2 d\Pi_3 d\Pi_4 e^{-\frac{E_1+E_2}{T}} (2\pi)^4 |\mathcal{M}_{1+2 \rightarrow 3+4}|^2 \delta^4(p_1+p_2-p_3-p_4).
\end{equation}
This equation takes into account both a process and its charge conjugate, resulting in the initial factor of 2.  It assumes the chemical potentials $\mu_1$ and $\mu_2$ are much less than the temperature $T$.  Here, 
$d\Pi_i \equiv \frac{d^3 p_i}{(2\pi)^3 2 E_i}$, and $|\mathcal{M}|^2$ is the squared matrix element summed over initial and final spins, polarizations, and gauge group indices.

We now turn to the calculation of the relevant matrix elements. One process that contributes is $\ell_i+W \rightarrow N^{\dagger}_j+H^{\dagger}_u$.  Here, $i$ and $j$ are generation indices. There are two other permutations of this process that result from swapping the $H^{\dagger}_u$ with one of the initial particles.  For each of these three permutations, there are six possible processes once we account for the contribution of superpartners.  For example, in addition to the original process, we have $\ell_i+W \rightarrow \tilde{N}^{\dagger}_j+\tilde{H}^{\dagger}_u$, $\tilde{\ell}_i+W \rightarrow N^{\dagger}_j+\tilde{H}^{\dagger}_u$, $\ell_i+\tilde{W} \rightarrow \tilde{N}^{\dagger}_j+H^{\dagger}_u$, $\tilde{\ell}_i+\tilde{W} \rightarrow N^{\dagger}_j+H^{\dagger}_u$, and $\tilde{\ell}_i+\tilde{W} \rightarrow \tilde{N}^{\dagger}_j+\tilde{H}^{\dagger}_u$.  In principle, one could evaluate the amplitudes for all $3 \times 6=18$ of these processes separately.  However, different permutations of the initial- and final-state particles are related by crossing symmetry. Furthermore, we can relate many processes using the generators of the SUSY algebra to exchange particles with their superpartners. Taken together, these tools allow the computation of all 18 processes in terms of just one of the above processes. The results in the massless limit are shown in Table~\ref{tab:LWNH_Msqr}.

\begin{table}[t]
\begin{tabular}{c|c|c|c|}
\hline
\multicolumn{1}{|c|}{} & \multicolumn{1}{c|}{} & $2 \leftrightarrow 4$ & $1 \leftrightarrow 4$ \\ \Xhline{1.2pt}  
\multicolumn{1}{|c|}{$\ell_i+W \rightarrow N^{\dagger}_j+H^{\dagger}_u$} & \multicolumn{1}{c|}{$-3g_2^2|y_{ij}|^2\frac{u}{s}$} & $-3g_2^2|y_{ij}|^2\frac{s}{u}$ & $3g_2^2|y_{ij}|^2\frac{u}{t}$\\ \hline
\multicolumn{1}{|c|}{$\ell_i+W \rightarrow \tilde{N}^{\dagger}_j+\tilde{H}^{\dagger}_u$} & \multicolumn{1}{c|}{$-3g_2^2|y_{ij}|^2\frac{u^2}{st}$} & $3g_2^2|y_{ij}|^2\frac{s^2}{ut}$ & $-3g_2^2|y_{ij}|^2\frac{u^2}{st}$ \\ \hline
\multicolumn{1}{|c|}{$\tilde{\ell}_i+W \rightarrow N^{\dagger}_j+\tilde{H}^{\dagger}_u$} & \multicolumn{1}{c|}{$3g_2^2|y_{ij}|^2\frac{u}{t}$} & $-3g_2^2|y_{ij}|^2\frac{s}{t}$ & $-3g_2^2|y_{ij}|^2\frac{u}{s}$ \\ \hline
\multicolumn{1}{|c|}{$\ell_i+\tilde{W} \rightarrow \tilde{N}^{\dagger}_j+H^{\dagger}_u$} & \multicolumn{1}{c|}{$3g_2^2|y_{ij}|^2\frac{u}{t}$} & $-3g_2^2|y_{ij}|^2\frac{s}{t}$ & $-3g_2^2|y_{ij}|^2\frac{u}{s}$\\ \hline
\multicolumn{1}{|c|}{$\tilde{\ell}_i+\tilde{W} \rightarrow N^{\dagger}_j+H^{\dagger}_u$} & \multicolumn{1}{c|}{$-3g_2^2|y_{ij}|^2\frac{u^2}{st}$} & $3g_2^2|y_{ij}|^2\frac{s^2}{ut}$ & $-3g_2^2|y_{ij}|^2\frac{u^2}{st}$ \\ \hline
\multicolumn{1}{|c|}{$\tilde{\ell}_i+\tilde{W} \rightarrow \tilde{N}^{\dagger}_j+\tilde{H}^{\dagger}_u$} & \multicolumn{1}{c|}{$-3g_2^2|y_{ij}|^2\frac{u}{s}$} & $-3g_2^2|y_{ij}|^2\frac{s}{u}$ & $3g_2^2|y_{ij}|^2\frac{u}{t}$ \\ \hline
\multicolumn{1}{|c|}{Sum} & \multicolumn{1}{c|}{$12g_2^2|y_{ij}|^2\frac{u}{t}$} & $12g_2^2|y_{ij}|^2\frac{s^2}{tu}$ & $12g_2^2|y_{ij}|^2\frac{u}{t}$ \\ \hline
\end{tabular}

\caption{Squared amplitude, $|\mathcal{M}|^2$, summed over initial and final spins, polarizations, and gauge group indices for all permutations of the process $\ell_i+W \rightarrow N^{\dagger}_j+H^{\dagger}_u$, in the massless limit.  Each row corresponds to a different permutation of superpartners, and each column corresponds to a different permutation of initial- and final-state particles.  Specifically, the column labeled $2 \leftrightarrow 4$ interchanges the $W$ or $\tilde{W}$ with the $H^{\dagger}_u$ or $\tilde{H}^{\dagger}_u$, and the column labeled $1 \leftrightarrow 4$ interchanges the $\ell_i$ or $\tilde{\ell}_i$ with the $H^{\dagger}_u$ or $\tilde{H}^{\dagger}_u$.  The final row shows the sum of all the superpartner permutations for one permutation of initial and final state particles.  This sum is useful if one assumes that particles have the same chemical potential as their superpartners, as is the case if chiral symmetry breaking interactions through the gaugino mass are in equilibrium \cite{Barnes:2022ren}.  Here, $g_2$ is the Standard Model $SU(2)_L$ gauge coupling, $y_{ij}$ is the Yukawa coupling between $\ell_i$ and $N_j$, and $s$, $t$, and $u$ are Mandelstam variables.}
\label{tab:LWNH_Msqr}
\end{table}

Another process that contributes is $\ell_i+t \rightarrow N^{\dagger}_j+q_3$.  Here, $t$ is the right-handed component of the top quark, and $q_3$ is the quark doublet in the third generation.  Again, there are two more permutations from switching initial and final state particles, and five more from switching particles with their superpartners: $\tilde{\ell}_i+t \rightarrow N^{\dagger}_j+\tilde{q}_3$, $\ell_i+t \rightarrow \tilde{N}^{\dagger}_j+\tilde{q}_3$, $\ell_i+\tilde{t} \rightarrow \tilde{N}^{\dagger}_j+q_3$, $\tilde{\ell}_i+\tilde{t} \rightarrow N^{\dagger}_j+q_3$, and $\tilde{\ell}_i+\tilde{t} \rightarrow \tilde{N}^{\dagger}_j+\tilde{q}_3$.  Just as before, we can use crossing symmetry and supersymmetry to relate the 18 relevant amplitudes.   The results are shown in the massless limit in Table ~\ref{tab:LtNQ_Msqr}.  Finally, there is the charge conjugate of every process mentioned so far.

\begin{table}[t]
\begin{tabular}{c|c|c|c|}
\hline
\multicolumn{1}{|c|}{} & \multicolumn{1}{c|}{} & $2 \leftrightarrow 4$ & $1 \leftrightarrow 4$ \\ \Xhline{1.2pt}  
\multicolumn{1}{|c|}{$\ell_i+t \rightarrow N^{\dagger}_j+q_3$} & \multicolumn{1}{c|}{$6y_t^2|y_{ij}|^2$} & $6y_t^2|y_{ij}|^2$ & $6y_t^2|y_{ij}|^2$\\ \hline
\multicolumn{1}{|c|}{$\tilde{\ell}_i+t \rightarrow N^{\dagger}_j+\tilde{q}_3$} & \multicolumn{1}{c|}{$-6y_t^2|y_{ij}|^2\frac{s}{t}$} & $6y_t^2|y_{ij}|^2\frac{u}{t}$ & $-6y_t^2|y_{ij}|^2\frac{t}{s}$ \\ \hline
\multicolumn{1}{|c|}{$\ell_i+t \rightarrow \tilde{N}^{\dagger}_j+\tilde{q}_3$} & \multicolumn{1}{c|}{$6y_t^2|y_{ij}|^2\frac{u}{t}$} & $-6y_t^2|y_{ij}|^2\frac{s}{t}$ & $-6y_t^2|y_{ij}|^2\frac{u}{s}$ \\ \hline
\multicolumn{1}{|c|}{$\ell_i+\tilde{t} \rightarrow \tilde{N}^{\dagger}_j+q_3$} & \multicolumn{1}{c|}{$-6y_t^2|y_{ij}|^2\frac{s}{t}$} & $6y_t^2|y_{ij}|^2\frac{u}{t}$ & $-6y_t^2|y_{ij}|^2\frac{t}{s}$ \\ \hline
\multicolumn{1}{|c|}{$\tilde{\ell}_i+\tilde{t} \rightarrow N^{\dagger}_j+q_3$} & \multicolumn{1}{c|}{$6y_t^2|y_{ij}|^2\frac{u}{t}$} & $-6y_t^2|y_{ij}|^2\frac{s}{t}$ & $-6y_t^2|y_{ij}|^2\frac{u}{s}$ \\ \hline
\multicolumn{1}{|c|}{$\tilde{\ell}_i+\tilde{t} \rightarrow \tilde{N}^{\dagger}_j+\tilde{q}_3$} & \multicolumn{1}{c|}{$6y_t^2|y_{ij}|^2$} & $6y_t^2|y_{ij}|^2$ & $6y_t^2|y_{ij}|^2$ \\ \hline
\multicolumn{1}{|c|}{Sum} & \multicolumn{1}{c|}{$-24y_t^2|y_{ij}|^2\frac{s}{t}$} & $-24y_t^2|y_{ij}|^2\frac{s}{t}$ & $24y_t^2|y_{ij}|^2$ \\ \hline
\end{tabular}

\caption{Squared amplitude, $|\mathcal{M}|^2$, summed over initial and final spins, polarizations, and gauge group indices for all permutations of the process $\ell_i+t \rightarrow N^{\dagger}_j+q_3$, in the massless limit. Here, $y_t$ is the top quark Yukawa coupling.}
\label{tab:LtNQ_Msqr}
\end{table}

When the relevant amplitudes are integrated over the final state phase space, the Mandelstam variables $t$ and $u$ that appear in the denominators give rise to divergences at a scattering angle $\cos \theta = \pm 1$.  These divergences are regularized by the thermal mass of the particle in the propagator, which are $m^2 \sim 0.1T^2$, see \cite{Chung:2009cb} for precise values. 

Motivated by this expected  cutoff due to thermal effects, we provide a phenomenological cutoff to the angular integral as follows: 
\begin{equation}
    -1 + \delta < \cos \theta < 1- \delta,
\end{equation}
with the expected cutoff $\delta$ to be determined by the thermal mass in the propagator. We~take
\begin{equation}
    \delta = \frac{c}{\sqrt{s}/T + g}
\end{equation}
with $c=0.1$, and $g$ the electroweak coupling.   We have checked that variation of the precise form of this cutoff has a relatively mild impact on the rate of baryon asymmetry production, of order a few 10s of percent. 

The running of the neutrino Yukawa coupling $y_N$ is a relatively small effect, with $y_N^2$ changing from its EW scale value by $5-10\%$ in the parameter range of interest. The running of the top Yukawa coupling, relevant for process $\ell t \rightarrow q N$ (and those related by crossings and supersymmetry) is a more significant effect, with $y_{\rm top}^2$ at $T_{S}$ decreased by roughly $30\%$ from its electroweak value.

\begin{table}[t]
\begin{tabular}{c|c|c|c|}
\hline
\multicolumn{1}{|c|}{} & \multicolumn{1}{c|}{$|\mathcal{M}|^2$} & $\dot{n}_{B-L}$ \\ \Xhline{1.2pt}  
\multicolumn{1}{|c|}{$H_u \rightarrow \ell^{\dagger}_i + N^{\dagger}_j$} & \multicolumn{1}{c|}{$2|y_{ij}|^2(m_{H_u}^2-m_{\ell}^2)$} & $\frac{|y_{ij}|^2\mu_{H_u}T}{8\pi^3 m_{H_u}^2}(m_{H_u}^2-m_{\ell}^2)^2$ \\ \hline
\multicolumn{1}{|c|}{$\tilde{H}_u \rightarrow \ell^{\dagger}_i + \tilde{N}^{\dagger}_j$} & \multicolumn{1}{c|}{$2|y_{ij}|^2(m_{\tilde{H}_u}^2+m_{\ell}^2)$} & $\frac{|y_{ij}|^2\mu_{\tilde{H}_u}T}{8\pi^3 m_{\tilde{H}_u}^2}(m_{\tilde{H}_u}^4-m_{\ell}^4)$ \\ \hline
\multicolumn{1}{|c|}{$\tilde{H}_u \rightarrow \tilde{\ell}^{\dagger}_i + N^{\dagger}_j$} & \multicolumn{1}{c|}{$2|y_{ij}|^2(m_{\tilde{H}_u}^2-m_{\tilde{\ell}}^2)$} & $\frac{|y_{ij}|^2\mu_{\tilde{H}_u}T}{8\pi^3 m_{\tilde{H}_u}^2}(m_{\tilde{H}_u}^2-m_{\tilde{\ell}}^2)^2$ \\ \hline
\end{tabular}

\caption{Squared amplitude, $|\mathcal{M}|^2$, summed over initial and final spins, polarizations, and gauge group indices for all significant decays, as well as their contributions to $\dot{n}_{B-L}$.  In the computation of $\dot{n}_{B-L}$, it is assumed that the masses are well below the temperature.  The thermal masses $m_{H_u}^2$, $m_{\tilde{H}_u}^2$, $m_{\ell}^2$, and $m_{\tilde{\ell}}^2$ are discussed in the text.}
\label{tab:decay_Msqr}
\end{table}

There are three decay processes  $(1 \rightarrow 2 + 3)$ that can contribute: $H_u \rightarrow \ell^{\dagger}_i + N^{\dagger}_j$, $\tilde{H}_u \rightarrow \ell^{\dagger}_i + \tilde{N}^{\dagger}_j$, and $\tilde{H}_u \rightarrow \tilde{\ell}^{\dagger}_i + N^{\dagger}_j$.  There are also the charge conjugates of these processes.  For two-body decays, the contribution to the rate of $B-L$ production is:
\begin{equation}
\dot{n}_{B-L} \supset \frac{2\mu_1}{T} \int d\Pi_1 d\Pi_2 d\Pi_3 e^{-\frac{E_1}{T}} (2\pi)^4 |\mathcal{M}_{1 \rightarrow 2+3}|^2 \delta^4(p_1-p_2-p_3).
\end{equation}
This equation takes into account both a process and its charge conjugate, resulting in the overall factor of 2.  It assumes the chemical potential $\mu_1 \ll T$.  
Again, $|\mathcal{M}|^2$ is a squared amplitude summed over initial and final spins, polarizations, and gauge group indices, see Table \ref{tab:decay_Msqr}.  In the calculation of this decay amplitude, we keep both the thermal masses of the particles, but also use a modified version of the spinors following \cite{Weldon:1982bn}.    With this prescription we find good agreement with the ``tree-level" calculation of right-handed neutrino production in \cite{Besak:2012qm}.  This reference notes the potential importance of the resummation of soft gauge interactions on neutrino production these decays, finding that these gauge interactions can give an enhancement of roughly a factor of 3 for RH neutrino production in the Standard Model at the temperatures of interest $T\sim 10^{6}$ GeV.  Even with this large enhancement, decays are still somewhat subdominant to $2\rightarrow 2$ scattering.  In our case of interest, there are superpartners in the bath. 
 The MSSM Debye mass is somewhat larger than the SM one, which should lead to increased screening and a somewhat reduced enhancement.  For concreteness, we set this enhancement to a factor of 2 relative to our ``tree calculation," a more precise calculation would require a detailed resummation along the lines of \cite{Besak:2012qm}. 

After taking into account the contributions from both scattering and decays, 
one can solve for the chemical potentials.  This can be done using the formalism described in the Appendix of Ref.~\cite{Barnes:2022ren}. At the temperatures of interest, all MSSM interactions are expected to be in equilibrium. The translation between the chemical potentials and $\dot{\theta}$ is shown in~Table~\ref{tab:chempot}.

\begin{table}[!t]
    \centering
    \begin{tabular}{c|c|c}
         &  DFSZ & KSVZ  \\
         \hline
$\mu_{\ell_{i}}$ & $\frac{59}{474} \frac{n}{N_{\rm DW}}$    & $\frac{1}{237}(12 c_g -25 c_W)$ \\
$\mu_{H_{u}}$ &   $-\frac{143}{158} \frac{n}{N_{\rm DW}} $   & $-\frac{1}{79} (9 c_g + c_W )$\\
$\mu_{q_{3}}$ &  $\frac{11}{158} \frac{n}{N_{\rm DW}}$   & $-\frac{2}{237} (2 c_g + 9 c_W )$ \\
$\mu_{t^{c}}$ & $\frac{66}{79} \frac{n}{N_{\rm DW}}$   &  $\frac{1}{237} (31 c_g +21 c_W )$ \\
    \end{tabular}
    \caption{Relation of the chemical potentials to the axion model parameters in units of $\dot{\theta}$.}
    \label{tab:chempot}
\end{table}

\nocite{apsrev41Control}
\bibliographystyle{apsrev4-1}
\bibliography{refs}

\end{document}